\documentclass[fleqn,usenatbib]{mnras}

\usepackage{newtxtext,newtxmath}

\usepackage[T1]{fontenc}

\usepackage{enumerate}

\DeclareRobustCommand{\VAN}[3]{#2}
\let\VANthebibliography\thebibliography
\def\thebibliography{\DeclareRobustCommand{\VAN}[3]{##3}\VANthebibliography}

\usepackage{graphicx}	
\usepackage{amsmath}	

\title[MTNG -- Semi-analytic galaxies on the lightcone]
{The MillenniumTNG Project: Semi-analytic galaxy formation models on the past lightcone}

\author[M.~Barrera et al.]{%
Monica Barrera$^{1}$\thanks{E-mail: 
mabar@mpa-garching.mpg.de},
Volker Springel$^{1}$,
Simon White$^{1}$, 
C\'esar Hern\'andez-Aguayo$^{1,2}$,
Lars Hernquist$^{3}$, 
\newauthor
Carlos Frenk$^{4}$,
R\"udiger Pakmor$^{1}$,
Fulvio Ferlito$^{1}$,
Boryana Hadzhiyska$^{3}$,
Ana Maria Delgado$^{3}$, 
\newauthor
Rahul Kannan$^{3}$,
and Sownak Bose$^{4}$
\\
\\
$^{1}$Max-Planck-Institut f\"ur Astrophysik, Karl-Schwarzschild-Str. 1, D-85748, Garching, Germany\\
$^{2}$Excellence Cluster ORIGINS, Boltzmannstrasse 2, D-85748 Garching, Germany\\
$^{3}$Harvard-Smithsonian Center for Astrophysics, 60 Garden Street, Cambridge, MA 02138, USA\\
$^{4}$Institute for Computational Cosmology, Department of Physics, Durham University, South Road, Durham, DH1 3LE, UK
}

\date{Accepted 2023 August 24. Received 2023 July 14; in original form 2022 October 24}

\pubyear{2023}

\begin{document}
\label{firstpage}
\pagerange{\pageref{firstpage}--\pageref{lastpage}}
\maketitle

\begin{abstract}
Upcoming large galaxy surveys will subject the standard cosmological model, $\Lambda$CDM, to new precision tests. These can be tightened considerably if theoretical models of galaxy formation are available that can predict galaxy clustering and galaxy-galaxy lensing on the full range of measurable scales, throughout volumes as large as those of the surveys, and with sufficient flexibility that uncertain aspects of the underlying astrophysics can be marginalised over. This, in particular, requires mock galaxy catalogues in large cosmological volumes that can be directly compared to observation, and can be optimised empirically by Monte Carlo Markov Chains or other similar schemes, thus eliminating or estimating parameters related to galaxy formation when constraining cosmology. Semi-analytic galaxy formation methods implemented on top of cosmological dark matter simulations offer a computationally efficient approach to construct physically based and flexibly parametrised galaxy formation models, and as such they are more potent than still faster, but purely empirical models. Here we introduce an updated methodology for the semi-analytic {\small L-GALAXIES} code, allowing it to be applied to simulations of the new MillenniumTNG project, producing galaxies directly on fully continuous past lightcones, potentially over the full sky, out to high redshift, and for all galaxies more massive than $\sim 10^8\,{\rm M}_\odot$. We investigate the numerical  convergence of the resulting predictions, and study the projected galaxy clustering signals of different samples. The new methodology can be viewed as an important step towards more faithful forward-modelling of observational data, helping to reduce systematic distortions in the comparison of theory to observations.
\end{abstract}

\begin{keywords}
galaxies: formation -– galaxies: evolution –- methods: semi-analytical
\end{keywords}

\section{Introduction}

Current and forthcoming observational programs, such as DESI or Euclid, target survey volumes of unprecedented size, with the number of observed galaxies reaching into the billions. This enormous size stresses the need for constructing equally large theoretical mock catalogues, because only then the full constraining power of the data can be harvested. There are a number of different methods that in principal allow the production of big enough mock galaxy surveys for validating and testing our galaxy formation theories. Unfortunately, the most direct approach -- hydrodynamical cosmological simulations -- requires too much computational power to cover the necessary target volumes or to vary uncertain parameters over their plausible ranges \citep[see][for a review]{Vogelsberger2020}. 

Alternatively, dark matter only simulations can be used to construct more approximate semi-analytic galaxy formation models. While they still follow the hierarchical build-up of structures quite faithfully, they treat baryonic physics very coarsely and neglect its impact on matter clustering, thus they have  higher systematic uncertainties than the hydrodynamical models. There are also computationally still less expensive options, in the form of empirical approaches, such as halo occupation distribution \citep[e.g.][HOD]{Berlind2002} subhalo abundance matching \citep[e.g.][SHAM]{Conroy2006}, or empirical galaxy formation models \citep[e.g][]{Moster2013, Behroozi2019}. While such statistical approaches to the galaxy--halo connection \citep[see][for a review]{Wechsler2018} can be  useful, they do not fully enforce physical consistency when modelling galaxy formation, and this risks  weakening their overall constraining power.

In this work, we concentrate on semi-analytic models (SAMs), with the aim to apply them to a  new simulation suite, MillenniumTNG, in a form that gives the outputs higher fidelity and makes them more directly comparable to observations. In particular, we present a methodology that produces a fully continuous lightcone output of galaxies. This can, for example, be combined directly with  weak-gravitational lensing predictions produced in an equally continuous way by our simulation project, through high-resolution projections of the particle lightcone. Furthermore, our simulation set also includes a hydrodynamic, full physics simulation with the same initial conditions as one of our dark matter models. This allows direct comparison to the semi-analytic model, and thus for it to be to tested and further improved \citep{Ayromlou2021a}.

SAMs were originally conceived in seminal papers by \citet{White1978} and \citet{White1991}, and then became substantially more complex over the years, both by the adoption of refined physics \citep[e.g.][for black holes]{Kauffmann2000, Croton2006}, and by replacing random realisations of dark matter merger trees by trees directly measured from simulations \citep{Kauffmann1999}, initially only at the halo level, but eventually with all resolved dark matter substructures included \citep{Springel2001}. Over the past two and a half decades, semi-analytic models \citep[see][for reviews]{Baugh2006, Somerville2015} have been continuously refined and developed by many groups \citep[e.g.][]{Sommerville1999, Cole2000, Moncao2007, Somerville2008, Benson2012, Stevens2016, Cora2018, Lagos2018, Cattaneo2020, Gabrielpillai2022}. They have also been outfitted with techniques to create mock lightcone outputs \citep[e.g.][]{Blaizot2005, Kitzbichler2007, Merson2013, Sommerville2021, Yung2022, Yung2023} usually by suitably combining a set of outputs at discrete redshifts.

Only in recent years has serious competition to SAMs arisen for modelling galaxy formation physics throughout cosmological volumes, in the form of the first successful and moderately large-volume hydrodynamical simulations of galaxy formation, such as Illustris \citep{Vogelsberger2014}, EAGLE \citep{Schaye2015}, HorizonAGN \citep{Dubois2016}, IllustrisTNG \citep{Springel2018}, SIMBA \citep{Dave2019}, or Thesan \citep{Kannan2022t}. While these calculations provide a more accurate treatment especially of gas dynamics and galaxy structure, they are also much more computationally expensive and are still subject to similar fundamental uncertainties in modelling subgrid physics related to star formation and associated feedback processes. In addition, their computational cost restricts them to substantially smaller volumes.

In semi-analytic models, the baryonic physics of galaxy formation, such as radiative cooling, star formation, and associated feedback processes, is described in terms of simplified differential equations with different efficiency parameters. The latter are set through a calibration step using selected observational constraints. For example, the {\small L-GALAXIES} model, sometimes known as the `Munich model' often used the stellar mass function at a range of redshifts as constraints, with other galaxy properties such as clustering then being treated as predictions. SAMs have sometimes been criticised for their sizable number of free parameters, implying a degree of modelling freedom that might compromise the predictive power of the approach. It needs to be conceded, however, that hydrodynamical simulations are only moderately better in this respect, as they equally require numerous parameters for sub-grid prescriptions in need of calibration. Also, instead of tuning the free parameters in an ad-hoc way (in the old days done by trial and error, assisted by physical intuition), modern SAM approaches use systematic parameter searches, for example, by exploring the space of parameters using Monte Carlo Markov Chain (MCMC) methods \citep{Henriques2009}, which then can also inform about degeneracies and uncertainties of these parameters. Such MCMC optimisation is not feasible for hydrodynamical simulations.

In this work we develop a new version of the {\small L-GALAXIES}  semi-analytic model, starting from the version described in \citet{Henriques2015}, which in turn evolved via many intermediate versions and improvements \citep[e.g.][]{Guo2011, Guo2013, Henriques2013}  from code written by \citet{Springel2005} for the Millennium simulation. Our primary goal is to modernise the time integration methodology such that more accurate continuous outputting along the past lightcone becomes possible. Additionally, we have made the tracking of merger trees more accurate and robust, and we have substantially accelerated the code and modernised all parts of its infrastructure, both to facilitate applications to extremely large simulations, and to simplify future extensions and refinements of the physics model which is here adopted unchanged from \cite{Henriques2015}.

This study is part of a set of introductory papers for the new  MillenniumTNG project. \citet{Aguayo2022} give a detailed overview of the numerical simulations and an analysis of non-linear matter clustering and halo statistics, while \citet{Pakmor2022} present the large hydrodynamical simulation of the project and an analysis of its population of galaxy clusters. In \citet{Kannan2022}, we  investigate properties of very high redshift galaxies and compare them to the new observations made by JWST. \citet{Hadzhiyska2022a, Hadzhiyska2022b} present an analysis of HOD techniques and their shortcomings in light of galaxy assembly bias, while \citet{Delgado2022} study intrinsic alignments of galaxy and halos shapes. \citet{Contreras2022} introduce an inference methodology able to constrain cosmological parameters from galaxy clustering. Finally, \citet{Bose2022} consider  galaxy clustering for different colour-selected galaxy samples, while \citet{Ferlito2022} study weak-lensing convergence maps at very high resolution based on lightcone outputs of the simulations. 

The present paper is structured as follows. In Section~\ref{sec:sims} we describe the cosmological simulations of our MillenniumTNG project and detail their outputs used for this study. In Section~\ref{sec:lgalaxies}, we describe the {\small L-GALAXIES} semi-analytic model, and in particular, the changes we developed in order to make the model produce accurate and continuous lightcone outputs. We then turn to coding and convergence tests in Section~\ref{sec:convergence}, while in Section~\ref{sec:gallightcone} we use the galaxies on the lightcone obtained with the model for the two realisations of the MTNG simulation to study the projected two-point clustering signal. Finally, in Section~\ref{sec:conclusions} we summarise our findings and present our conclusions, in Appendix~\ref{appendix:lightconeclustering} we discuss cosmic variance effects for clustering measurements on the lightcone, and  in Appendix~\ref{appendix:speed} we discuss the speed of the new version of the semi-analytic code.

\section{Simulation set}
\label{sec:sims}

\subsection{The MillenniumTNG Project}

The MillenniumTNG project consists of several cosmological simulations of structure formation of the $\Lambda$CDM model, including pure dark matter simulations in a $500\,h^{-1}{\rm Mpc} \simeq 740\,{\rm Mpc} $ boxsize, a matching high-resolution hydrodynamical simulation, as well as several runs that additionally follow massive neutrinos as a small hot dark matter admixture. The overarching goal of the project is to link studies of galaxy formation and cosmic large-scale structure more closely in order to advance the theoretical understanding of this connection, which can ultimately be of help for carrying out precision tests of the cosmological model with the upcoming galaxy surveys.

Our main set of dark matter simulations uses a  $(740\,{\rm Mpc})^3$ volume, the same size as employed by the original Millennium simulation \citep{Springel2005} but at varying mass resolution reaching nearly an order of magnitude better. Our flag-ship hydrodynamical simulation uses the same large volume, and it is based on the physics model
\citep{Weinberger2017, Pillepich2018b} employed in the smaller IllustrisTNG simulations \citep{Nelson2018, Springel2018, Marinacci2018, Pillepich2018, Naiman2018, Pillepich2019, Nelson2019a, Nelson2019b}. Building upon the legacy of these two influential projects has motivated us to coin the name `MillenniumTNG' for our project, or MTNG for short. Following the notation of IllustrisTNG, we refer to our main runs as `MTNG740' and `MTNG740-DM', respectively.

Our  hydrodynamical simulation of galaxy formation expands on the important attribute of volume by nearly a factor of 15 compared to the previously leading model TNG300. While numerous other dark matter simulation projects exist in the literature with comparable or even larger particle number, for example MICE \citep{Fosalba2015}, MultiDark \citep{Klypin2016}, Uchuu \citep{Ishiyama2021}, BACCO \citep{Angulo2021} and AbacusSummit \citep{Maksimova2021}, and a few also feature even higher dark matter resolution than carried out here, for example Millennium-II \citep{Boylan-Kolchin2009} and Shin-Uchuu \citep{Ishiyama2021}, the combination of volume and resolution we reach in MTNG is still rare in the literature. Furthermore, we compute each of the dark matter models twice using a variance suppression technique \citep{Angulo2016}, which boosts the effective volume available for statistics on large scales. In addition,  we have augmented MillenniumTNG with still larger runs that evolve dark matter together with live massive neutrinos, going up to 1.1 trillion particles in a volume 68 times bigger than that of MTNG740-DM. We defer, however, the presentation of semi-analytic models for this extremely large simulation to forthcoming work, and focus here on introducing our new methodology using our dark matter simulation series in the standard box size of $740\,{\rm Mpc}$. The main parameters of the corresponding simulations are listed in Table~\ref{tab:resolutions}.

The computations have been carried out with the {\small GADGET-4} simulation code \citep{Springel2021}, apart from the hydrodynamical runs, which employed the moving-mesh code {\small AREPO} \citep{Springel2010}. A number of important improvements have been realised consistently in both codes compared to older versions of {\small GADGET} \citep{Springel2005b} and {\small AREPO}. Particularly relevant for the present work are better algorithms for identifying and tracking substructures, as well as the option of obtaining lightcone outputs while a simulation runs. For full details on the simulation set and the underlying numerical methodology, we refer the reader to our companion papers, particularly \citet{Aguayo2022}, as well as the code papers. In the following subsections, we will however briefly discuss the key aspects of halo finding and merger tree construction, as well as the lightcone outputting, as these are central for the analysis presented in this paper.

\subsection{Dark-matter only runs and merger trees}

Our dark matter simulations are based on initial conditions computed with $2^{\rm nd}$-order Lagrangian perturbation theory at redshift $z_{\rm init} = 63$. The cosmological model is the same as used for the IllustrisTNG simulations, and characterised by $\Omega_{\rm m} =\Omega_{\rm dm} + \Omega_{\rm b} = 0.3089$, $\Omega_{\rm b} = 0.0486$, $\Omega_\Lambda=0.6911$, and a Hubble constant $H_0 = 100\,h\, {\rm km\, s^{-1}Mpc^{-1}}$ with $h=0.6774$.  We use the `fixed-and-paired' technique of \citet{Angulo2016} to create two simulations at each given resolution that differ only by the sign of the imprinted linear density fluctuations. Furthermore, the amplitudes of all imprinted waves are set proportional to the square-root of the power spectrum at the corresponding wave vector instead of being drawn from a Rayleigh distribution. This means that the power in each mode is individually set to its expectation value, thereby reducing cosmic variance on large scales where only a few modes contribute, while on smaller scales the density fluctuation field resulting from the overlap of many modes is indistinguishable from standard realisations in terms of late time statistics. Additionally, averaging the results of the paired realisations eliminates leading order deviations from pure linear theory, so that on large scales the average of the paired simulations stays much closer to the expected cosmological mean than a normal realisation of the same volume would.

We have systematically varied the mass resolution by factors of eight to create a series of five different resolutions, ranging from a comparatively low resolution of 19.7 million particles up to 80.6 billion. This facilitates precise convergence tests, in particular also for the semi-analytic model, where this is less well understood than for the N-body particle simulations themselves. We identify halos and subhalos on-the-fly at a minimum of 265 snapshot times\footnote{A subset of the simulations has a couple of extra output times that augment the 265 regular output times present for all simulations.} that are spaced as follows:
\begin{itemize}
\item  $\Delta {\rm log}(a)= 0.0081\;\;$ for $\;\;0 \le z < 3\;\;\;$  (171 snapshots)
\item  $\Delta {\rm log}(a)= 0.0162\;\;$ for $\;\;3 \le z < 10\;\;$   (62 snapshots),
\item  $\Delta {\rm log}(a)= 0.0325\;\;$ for $\;\;10 \le z < 30\;\;$   (32 snapshots). 
\end{itemize}
The logarithmic intervals in expansion factor imply output spacings that correspond to fixed fractions of the current dynamical time of halos (and equivalently to the current Hubble time). The physical time between outputs varies between 116~Myr at $z=0$ to 25.8 Myr at $z=3$. It would shrink further towards higher redshift, reaching 1.22 Myr at $z=30$, if we would not have decided to reduce the output frequency by a factor of two for $3<z<10$, and by a further factor of two at even higher redshift, so that our finest temporal spacing at  $z=30$ is really 4.88~Myr. This coarser spacing at high redshift was adopted purely as a means to save computational resources, in particular disk storage, but also because there are fewer subhalos to track at high redshift, and in this regime it is arguably less critical to track their orbits within halos with high temporal accuracy. At each output time, we first run the friends-of-friends (FOF) group finding algorithm with a standard linking length of 0.2 times the mean particle spacing. Groups with a minimum particle number of 32 are retained and stored, as in most previous work with {\small L-GALAXIES}. While this particle number is too small for reliable measurements of internal properties of the smallest halos, such as density profile or shape, these quantities are presently 
not used in our semi-analytic model. We only use mass, position, velocity, spin parameter, and the maximum circular velocity of a (sub)halo. Furthermore, the expected slight excess of FOF halo counts close to the detection threshold \citep{Warren2006} is alleviated by processing all halos with an algorithm that identifies gravitationally bound subhalos within each group, filtering out spurious structures resulting from noise peaks. The unbinding approach is based on a classic self-potential binding check, but the recently suggested `boosted potential' of \citet{Stuecker2021} could be alternatively employed in the future to more naturally incorporate the effect of tidal fields. The improved substructure identification we use is based on the {\small SUBFIND-HBT} algorithm \citep{Han2018, Springel2021}, which in contrast to  previous versions of {\small SUBFIND} \citep{Springel2001} uses information from the subhalo catalogue at the previous output time. As shown in \citet[][their Fig.~36]{Springel2021}, the masses of subhalos are more accurately measured close to pericentre, improving the accuracy and robustness of tracking, and thus ultimately the quality of the merger trees constructed from the group-finder output.

For each subhalo, a variety of properties are automatically measured, such as the maximum rotation velocity $v_{\rm max}$, the radius at which this is attained $r_{\rm max}$, the velocity dispersion, the shape, the bulk velocity and the subhalo centre (taken as the position of the particle with minimum potential), the most-bound particle ID, etc. The largest bound subhalo in each FOF group is interpreted as the main background halo. Its centre is adopted as primary group centre, and is used to measure a number of masses defined through spherical apertures (these take always the full particle distribution into account, not just the gravitationally bound material). The most important of these spherical overdensity masses is $M_{200}$, the mass contained in a radius with overdensity of 200 relative to the critical density. In the neutrino runs, we have added a measurement of further subhalo properties, such as the environmental densities defined recently in \citet{Ayromlou2021b} to support an improved model for ram pressure stripping.

\begin{figure*}
\resizebox{15cm}{!}{\includegraphics{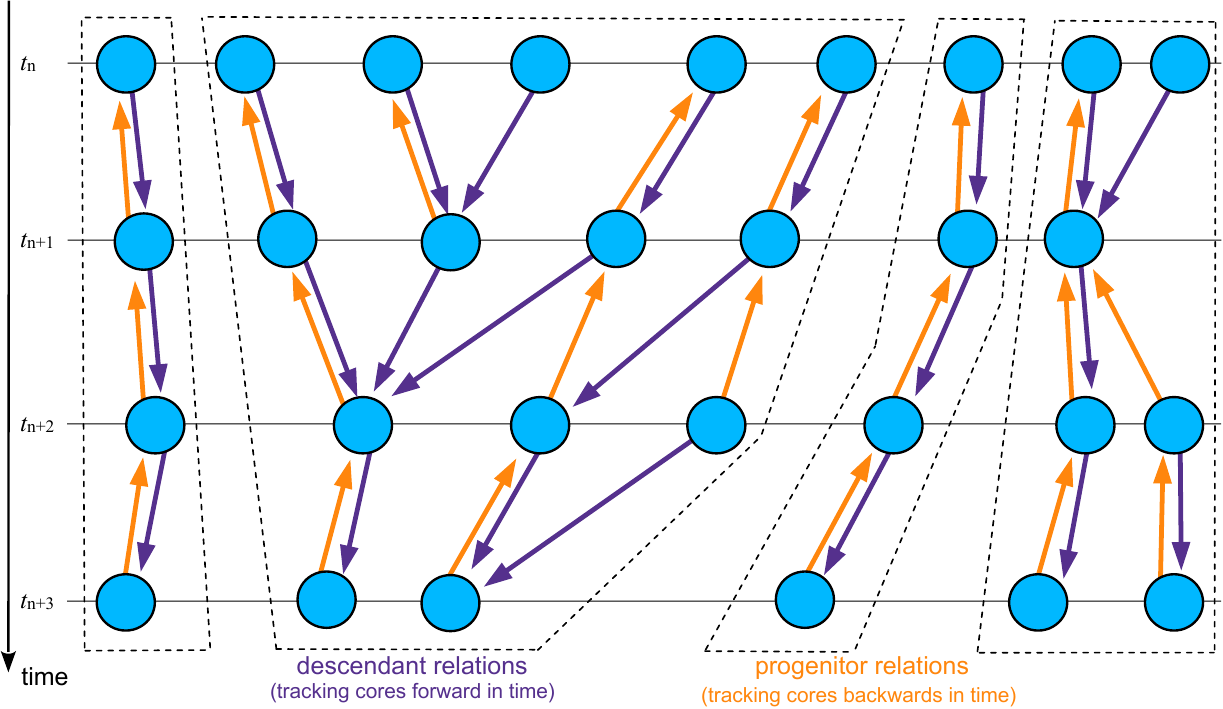}}
\caption{Sketch illustrating the primary temporal subhalo links used by our simulation code to define the merger tree used by the semi-analytic code. Fundamentally, we build the galaxies on merger trees composed of gravitationally bound subhalos that are tracked in time. The descendant and progenitor links between two subsequent times of group finding (e.g.~$t_{n+1}$ and $t_{n+2}$) are found on-the-fly during the N-body simulation itself. Besides these pointers, we also keep track of which subhalos are in the same FOF group (not shown in the sketch). When the simulation is finished, we identify the actual trees as those subsets of subhalos that are connected via at least one progenitor or descendant relationship, or through common membership in the same FOF group. The dashed lines indicate the four independent trees that would be identified in this particular example.
\label{fig:sketchtree}}
\end{figure*}

Our simulations do not store actual particle data for (most of) the defined snapshot times\footnote{Except that we have still done this for 10 selected snapshots to support other types of analysis that look at time-slices at the level of raw data. For the purposes of this paper, these outputs, each weighing 2.8 TB for one of our $4320^3$ runs, are not needed.}, i.e.~the information which particles make up a subhalo is not saved on disk in order to eliminate the taxing storage cost this would entail. Instead, the code links, already during runtime, the subhalos of a newly created snapshot catalogue with the most recent catalogue that was determined at a previous time. This is done by considering the 20 most bound particles in each subhalo and looking up in which gravitationally bound subhalo they are found in the other snapshot, identifying in this way the most likely descendant of a subhalo when one carries out this search in the forward direction. Likewise, the most likely progenitor of a subhalo is determined by carrying out the search in the backwards direction \citep[see][for a detailed description of the  procedure]{Springel2021}. These links are stored, and subsequently used (once the simulation has finished) to identify the full merger trees of a simulation.  A schematic sketch of the logical tree structure is sketched in Figure~\ref{fig:sketchtree}. Note that while the progenitor and descendant pointers typically simply occur in pairs that are opposite to each other, this is not the case when two or more subhalos merge. Then a subhalo may have multiple progenitors pointing to it. Only this case was treated in previous versions of our formalism, but our new merger trees can also account for situations where multiple descendants point to the same progenitor subhalo. This can happen, for example, during a (grazing) collision of subhalos that come apart again. It is a rare occurrence, however, only 0.24 percent of the subhalos in our trees are identified to be a potential progenitor of more than one subhalo. Notice that a satellite galaxy that comes out on the other side of a halo, a so-called splashback galaxy \citep{Diemer2021}, does not typically manifest itself through such a feature in our merger trees because usually we can track a splashback galaxy  unambiguously attached to its own subhalo.

Single trees are defined as follows in the merger tree building: Two subhalos are in the same tree if they are linked either by a descendant or by a progenitor pointer. They are also guaranteed to be in the same tree if they are member of the same FOF group. Finally, if any two subhalos have the same particle ID as their most-bound particle, they are also guaranteed to be in the same tree. These three equivalence class relations induce a grouping of the subhalos into disjoint sets (the trees) that guarantee that our semi-analytic galaxy formation model can be executed on each tree without requiring any extra information from a subhalo outside of the tree. A consequence is that trees can be processed in parallel if desired, with no side-effects on each other. Note, however, that there is not necessarily a single FOF group at $z=0$ for every tree . For example, if a thin particle bridge happens to link two groups at some earlier time (so that they form a single FOF halo at this time), for example in a grazing collision, all their descendants will be in the same tree structure, even if this involves having two (or more) disjoint FOF halos at $z=0$.

Compared with the Millennium project we have about four times as many output times, yielding a better time resolution of the merger tree. Also, the tracking of subhalos is more accurate and robust, and the addition of progenitor pointers allows recovery from edge cases for which proper tracking would otherwise be lost. We nevertheless retain the concept of `orphan galaxies', which refers to galaxies that are hosted by subhalos that are destroyed by gravitational tides before the galaxy has actually merged with the central galaxy. Including these objects significantly improves small-scale clustering predictions and the radial number density profiles of satellite galaxies in clusters \citep{Guo2011, Guo2013}. In order to keep some information about their spatial positions, we tag these galaxies with the ID of the most-bound particle at the last time the subhalo could still be identified, and then use the current coordinate of this particle as a proxy for the current galaxy position until it is finally predicted to merge with the central galaxy of its host halo. In order to have the phase-space coordinates of these particles available if needed at  future times (whether a certain ID's position will actually be needed, and for how long, depends on details of the semi-analytic model, such as its dynamical friction treatment), we actually store these particles in the form of `mini-snaphots' at snapshot times. The total number of particles accumulated in this way, i.e.~particles that have at some point been the most-bound particle of a gravitationally bound subhalo, reaches 3-4\% of all particles, which is a size that can still be managed well.

In Table~\ref{tab:resolutions}, we include some basic information about the number of halos, subhalos, and trees, as well as the total cumulative number of subhalos in the full forest of trees of each of the resolution levels. We shall refer to the individual runs with names such as MTNG740-DM-1-A, where the first number encodes the box size in Mpc, and the `A' refers to variant A of the pair of two simulations run at this resolution level 1. The letter B labels the mirrored realisation \citep[see][for a table of all simulations of the MTNG project]{Aguayo2022}.

\begin{table*}
	\centering
	\begin{tabular}{lcccccccc}
		\hline
		Run names (A|B), all & Particles & $m_{\rm DM}$ & Softening & \# FOF groups & \# subhalos & total \# of\\
		$L_{\rm box} = 500\,h^{-1}{\rm Mpc}$         &  & [$h^{-1}\mathrm{M}_\odot$]  & [$h^{-1}\mathrm{kpc}]$ & at $z=0$ & at $z=0$ &subhalos in trees\\
		\hline
		MTNG740-DM-5 & $270^3$ & $5.443\times 10^{11}$ & 40 & $38115\;|\;38108$  & $43034\;|\;42932$  & $3.250\times 10^{6}\;|\;3.217\times 10^{6}$  \\
		MTNG740-DM-4 & $540^3$ & $6.804\times 10^{10}$ & 20 & $279010\;|\;284219$  & $334755\;|\;340269$  & $3.775\times 10^{7}\;|\;3.890\times 10^{7}$  \\
		MTNG740-DM-3 & $1080^3$ & $8.505\times 10^{9}$ & 10 & $1.781\times 10^{6}\;|\;1.818\times 10^{6}$  & $2.215\times 10^{6}\;|\;2.262\times 10^{6}$  & $3.421\times 10^{8}\;|\;3.509\times 10^{8}$  \\
		MTNG740-DM-2 & $2160^3$ & $1.063\times 10^{9}$ & 5.0 & $1.127\times 10^{7}\;|\;1.146\times 10^{7}$  & $1.420\times 10^{7}\;|\;1.443\times 10^{7}$  & $2.736\times 10^{9}\;|\;2.785\times 10^{9}$  \\
          MTNG740-DM-1 & $4320^3$ & $1.329\times 10^{8}$ & 2.5 & $7.317\times 10^{7}\;|\;7.415\times 10^{7}$  & $9.135\times 10^{7}\;|\;9.266\times 10^{7}$  & $2.059\times 10^{10}\;|\;2.085\times 10^{10}$  \\
		\hline
	\end{tabular}
    \caption{Numerical parameters of the primary dark matter runs of the MillenniumMTNG project as analyzed in this work. These simulations have been carried out at five different resolutions in a periodic box $500\,h^{-1}{\rm Mpc} = 740\, {\rm Mpc}$ on side, and in two realisations A and B each. We list the symbolic run name, the particle number, the mass resolution, the gravitational softening length, the number of FOF groups at redshift $z=0$, the number of gravitationally bound subhalos at $z=0$, and the total cumulative number of subhalos in the merger trees. For the latter three quantities, we give the numbers for the A and B realisations separately.}
    \label{tab:resolutions}
\end{table*}

\subsection{Lightcone output}

In our MTNG simulations, we have produced several full particle lightcone outputs with different geometries \citep[see][]{Aguayo2022}. This data consists of the phase-space information of simulation particles at the instant they cross the past backwards lightcone of a fiducial observer placed into the simulation box\footnote{The observer position is simply the origin in our simulation box. Note that this point is not special in any way due to the periodic boundary conditions of the models.}. To realize this, the code checks during each particle timestep whether the particle crosses the lightcone during this step, and if so, the intersection is computed and stored. We produce such particle lightcones over the whole sky out to redshift $z=0.4$, for an octant of the sky out to $z=1.5$, and for a square-shaped `pencil-beam' with 10 degrees on a side out to redshift $z=5$. Doing a full-sky output out to a redshift as high as $z=5$ would produce a prohibitively large data volume, this is why we restrict ourselves to much narrower solid angles when going deeper. We do, however, internally construct the full particle lightcone out to $z=5$, but we project it right away into mass-shell projections on fine two-dimensional maps that can be used for weak-lensing studies \citep{Ferlito2022}.

Furthermore, we actually do store a subset of the particles of the fiducial full-sky lightcone out to $z=5$, but we only include particles that have been a most-bound particle of a subhalo at some point in the past. These particles can be used in our semi-analytic machinery to accurately reconstruct the time and location when {\em galaxies} cross the lightcone, because they are tracked by formerly most-bound particles in our approach (see below for more details). Still, the data volume accumulated in this way is substantial. For example, the $z=5$ full-sky most-bound particle lightcone of the MTNG740-DM-1-A simulation contains about $6.54 \times 10^{12}$ particle entries. Efficiently finding the right particle from this data set during a semi-analytic galaxy formation computation requires efficient storage and sorting/indexing approaches. One of the methods we use for this is to sort the lightcone data in a preprocessing step according to the tree it belongs to, which is a feature built in to the {\small GADGET-4} code for this purpose.

\section{Updates to  L-Galaxies}
\label{sec:lgalaxies}

As a starting point of our work we have used the stand-alone version of the {\small L-GALAXIES} code as
described in \citet{Henriques2015} (Hen15) and made publicly available by these authors\footnote{\url{https://lgalaxiespublicrelease.github.io}}. A detailed description of the physics model and its parametrization can be found in their supplementary material. We have kept the physical model of \citet{Henriques2015}   deliberately unchanged for the most part for the purposes of this work, apart from minor details\footnote{For example, we set the parameter {\tt SfrColdCrit} to 0 after finding no relevant changes compared to results where this constraint is imposed.}. However, we have very substantially modified the code at a technical level, primarily to improve its time integration schemes in order to facilitate continuous lightcone outputting of galaxy properties. Secondary goals of the changes have been to modernize the code architecture, to reduce its memory footprint, to make it more flexible in adjusting to differences in output spacing, to take advantage of additional features present in the new generation of merger trees we use, and finally, to move all I/O to the flexible and modern HDF5 scientific data format. In fact, to realize the latter point in an easy fashion, we ended up integrating {\small L-GALAXIES} as a postprocessing module into the {\small GADGET-4} simulation code, so that both codes can use the same C++ classes for organizing the I/O, for memory handling, and other functionality. The codes still remain logically distinct, however. The associated clean-up and partial rewrite  of the code-base of  {\small L-GALAXIES}  in the C++ language has led to a leaner and more easily extensible code.

In the following subsections, we will discuss in detail the most important changes relating to the time integration, the handling of `orphans' and galaxy orbits in general, as well as the treatment of the photometry. We note that a number of these changes and improvements were prompted by the new lightcone outputting functionality that we have realized. Previously, {\small L-GALAXIES} was essentially alternating between two discrete operations, an updating of the galaxy positions with the group catalogue information of a new snapshot time, and then evolving the equations describing the galaxy formation physics over the time interval between two snapshots using a number of small timesteps.

Because an output of galaxy properties only occurred at the snapshot times, this scheme was sufficient because both operations were always completed (i.e.~synchronized) at the output times. For continuous outputting in time, as needed for accurate lightcones, a number of subtle issues arise, most importantly the danger of introducing detectable ``discontinuities'' in galaxy properties along the redshift direction of lightcone outputs. For example, repositioning a galaxy at certain instants in time (when new subhalo catalogue information is introduced) to a new subhalo coordinate would appear as a sudden `teleportation' of a galaxy. Similarly, updating halo masses at discrete times would introduce discontinuous changes in the cooling and thus star formation rate of galaxies. The problem is not really that such jumps occur but rather that they occur for all galaxies at the same redshift, and thus at the same comoving distance (since the group and subhalo catalogues in the merger tree are computed at discrete times). This is undesirable. 

\begin{figure}
 \resizebox{8cm}{!}{\includegraphics{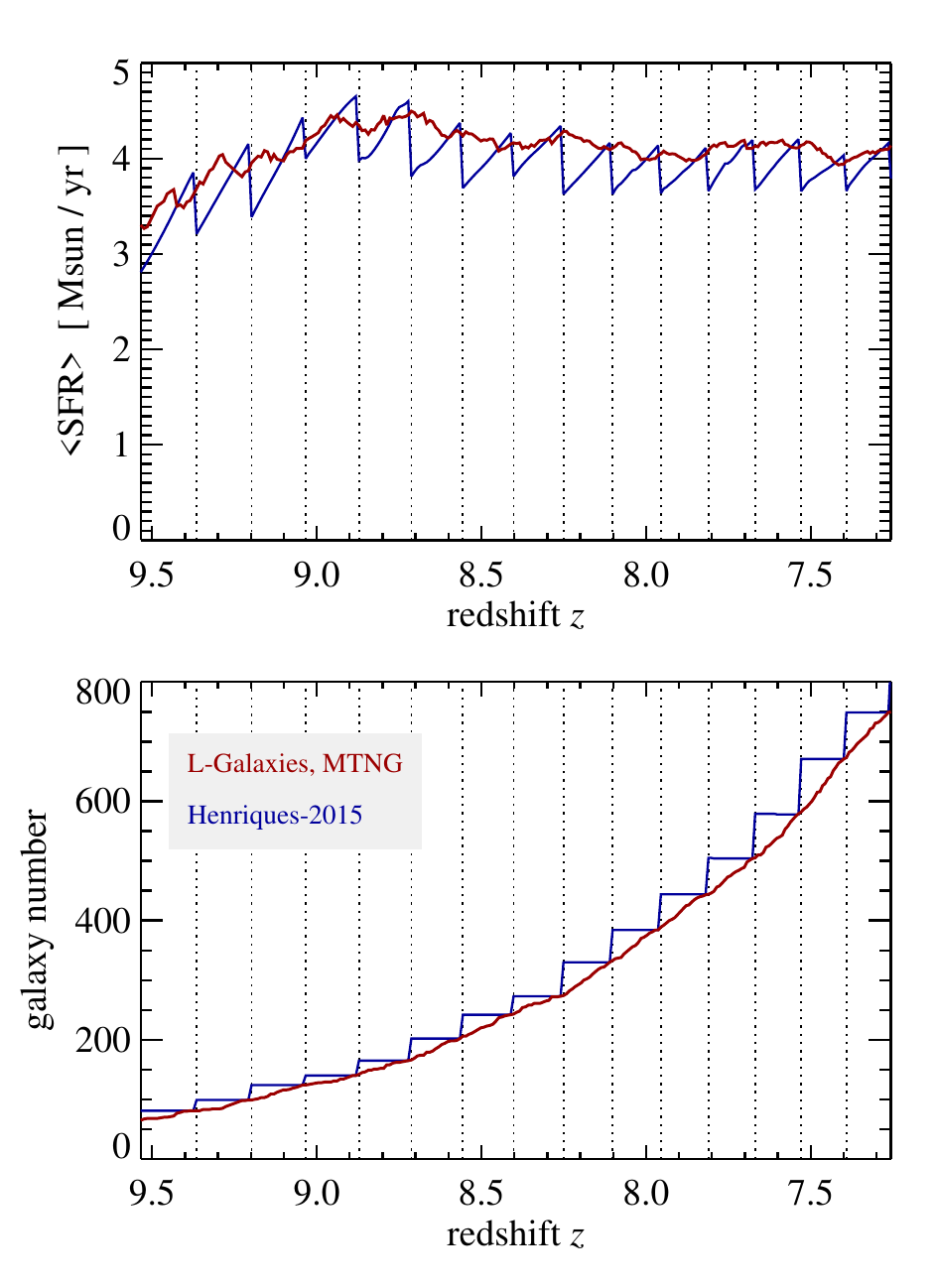}}
 \caption{The top panel shows the time evolution of the average star formation rate per galaxy at high redshift,
   in a typical large merger tree taken from the MTNG740-DM-2-A simulation, computed with the {\small L-GALAXIES} code. The bottom panel gives the evolution of the absolute number of galaxies that are tracked in this tree. We compare results for our new version of the semi-analytic code (red solid lines) with those for
  the older methodology of \citet{Henriques2015}, drawn as blue lines. This older version of {\small L-GALAXIES} created new galaxies always exactly at snapshot times (marked by the dotted vertical lines), when a new group catalogue becomes available. As a result, neither the galaxy number density nor the SFR density evolve continuously in time, but rather show sawtooth-like discontinuities that can induce faint spurious features in lightcone outputs. In our improved code, we randomize the birth time of new galaxies between two snapshot times in order to eliminate this effect.
   }
    \label{fig:inset}
\end{figure}

An example of such an effect is illustrated in Fig.~\ref{fig:inset}, where we show in the top panel the time evolution of the average star formation rate per galaxy at high redshift, in a typical large merger tree taken from the MTNG740-DM-2-A simulation. The bottom panel gives the evolution of the absolute number of galaxies that are tracked in this tree. The older methodology of \citet{Henriques2015} always creates new galaxies exactly at snapshot times (marked by the dotted vertical lines), when a new group catalogue becomes available. As a result, neither the galaxy number density nor the mean star formation rate evolve continuously in time, but rather show sawtooth-like discontinuities which can induce faint spurious features in lightcone outputs in the redshift direction. In our new improved code, this particular effect is eliminated by randomizing the birth time of new galaxies between two snapshot times.

A related problem concerns the photometry of galaxies. Previously, {\small L-GALAXIES} would already know the desired output redshifts when the galaxies were evolved in time. This made it possible to anticipate for any amount of newly formed stars how old the corresponding stellar population would be at the desired output times, and thus to integrate up the observed luminosity with the help of a spectrophotometric model taking the corresponding age differences into account. For a lightcone output, this scheme no longer works. Not only is the instant of lightcone crossing for a particular galaxy unknown as it evolves at higher redshift, also it may cross the lightcone at multiple different times once the periodic replication of the simulation box is taken into account. These issues can be resolved, however, if every galaxy keeps a sufficiently detailed record of its own star formation history, as described by \citet{Shamshiri2015}. We make use of the same idea here, but provide a new technical implementation that we describe in a dedicated subsection below.

\begin{figure*}
\resizebox{15cm}{!}{\includegraphics{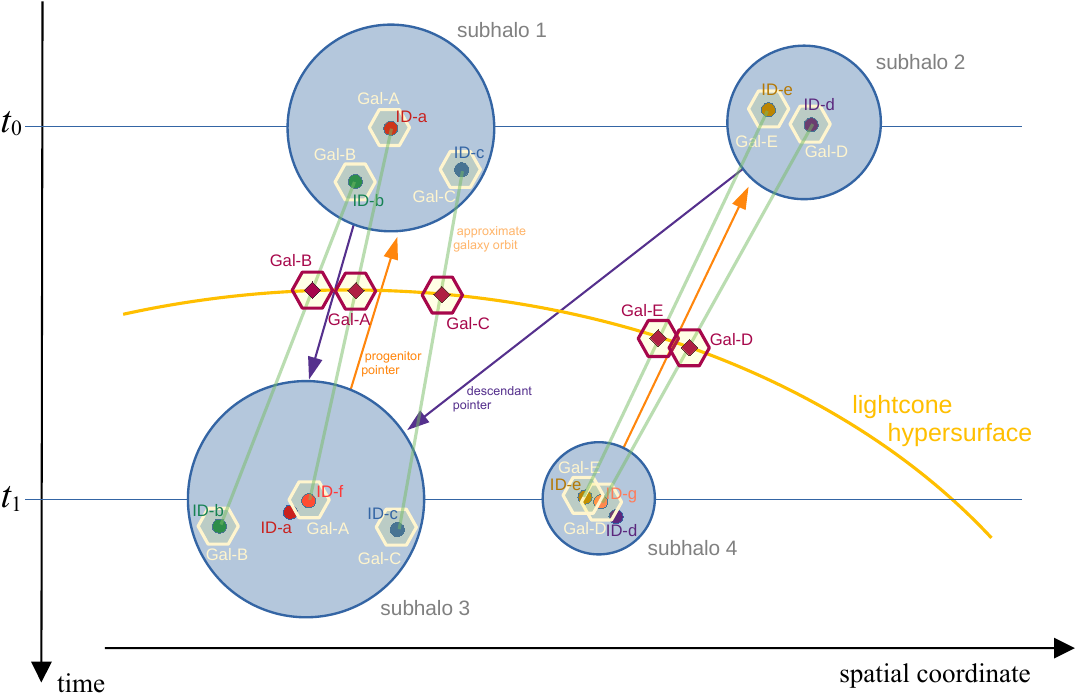}}
\caption{Sketch illustrating the time integration scheme of the semi-analytic code between two subsequent times $t_0$} and $t_1$ in the merger tree. The sketch considers 5 galaxies marked with hexagons and labelled Gal-A to Gal-E. Each of the galaxies is associated with a simulation particle ID, labelled ID-a to ID-e. For the galaxies in subhalo 1, only subhalo 3 is a possible new site at time $t_1$, while for the galaxies in subhalo 2, both subhalos at time $t_1$ are possible, due to the additional progenitor link pointing from subhalo 4 to subhalo 2. Galaxies D and E are ultimately assigned to subhalo~4, because the spatial distance of their particle IDs to the (new) particle ID tracking the center of subhalo 4, ID-g, is smaller than the distance to the center of subhalo 3, ID-f. In subhalo 3, galaxy A is selected as central galaxy and associated with a new particle ID, namely ID-f. The other two galaxies stay satellites and retain their particle IDs for tracking. In subhalo 4, galaxy D is selected as central galaxy, with its coordinate now being given by particle ID-g, the center of the corresponding subhalo. Note that particles ID-a and ID-b no longer represent galaxies at time $t_1$. Green straight lines mark linear orbit approximations between the two times $t_0$ and $t_1$ for the new galaxy positions. At their intersections with the lightcone, we obtain interpolated galaxy coordinates (red hexagons) together with galaxy properties integrated up to the corresponding times. 
\label{fig:sketchsam}
\end{figure*}

\begin{figure*}
\resizebox{8.5cm}{!}{\includegraphics{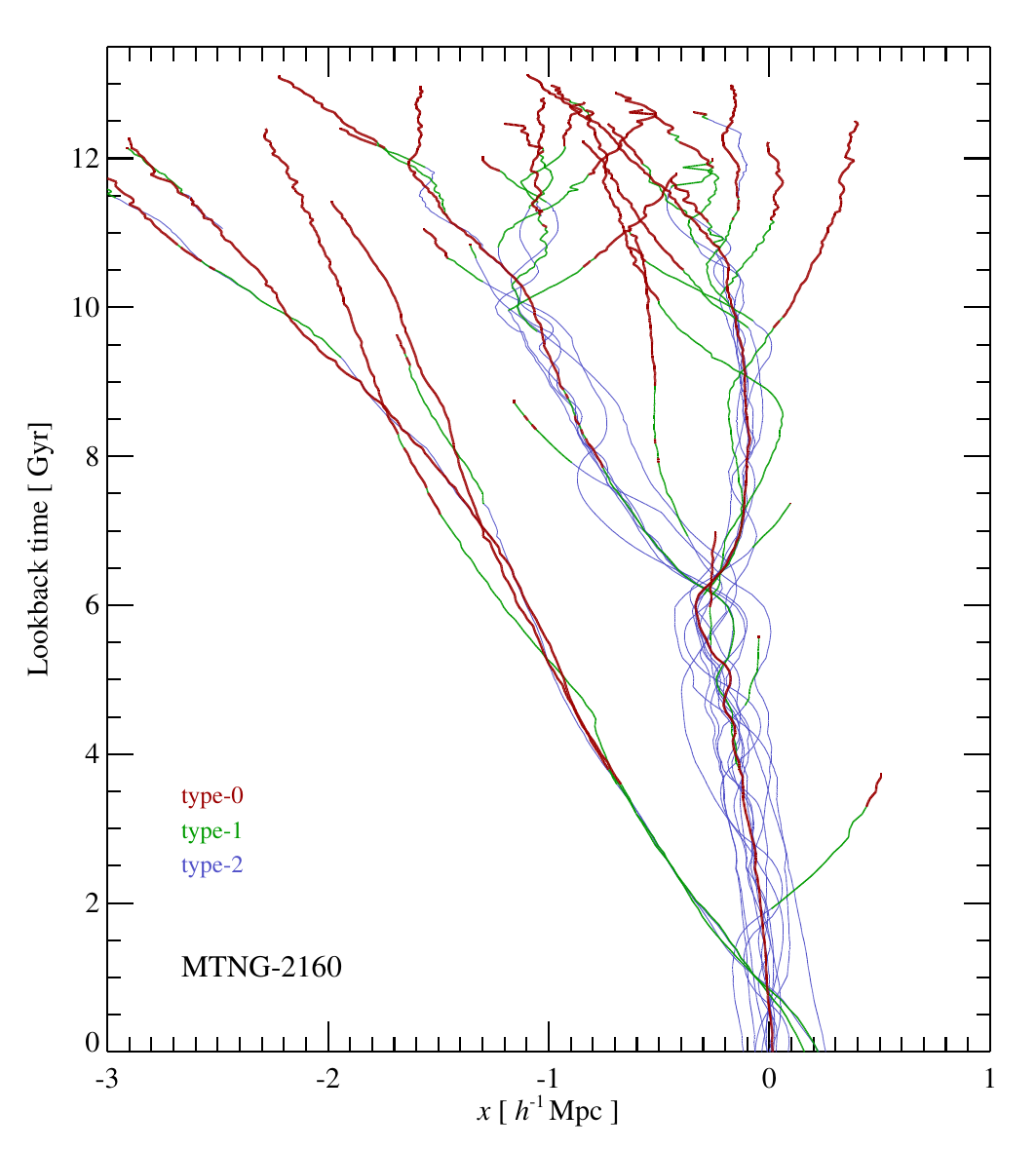}}%
\resizebox{8.5cm}{!}{\includegraphics{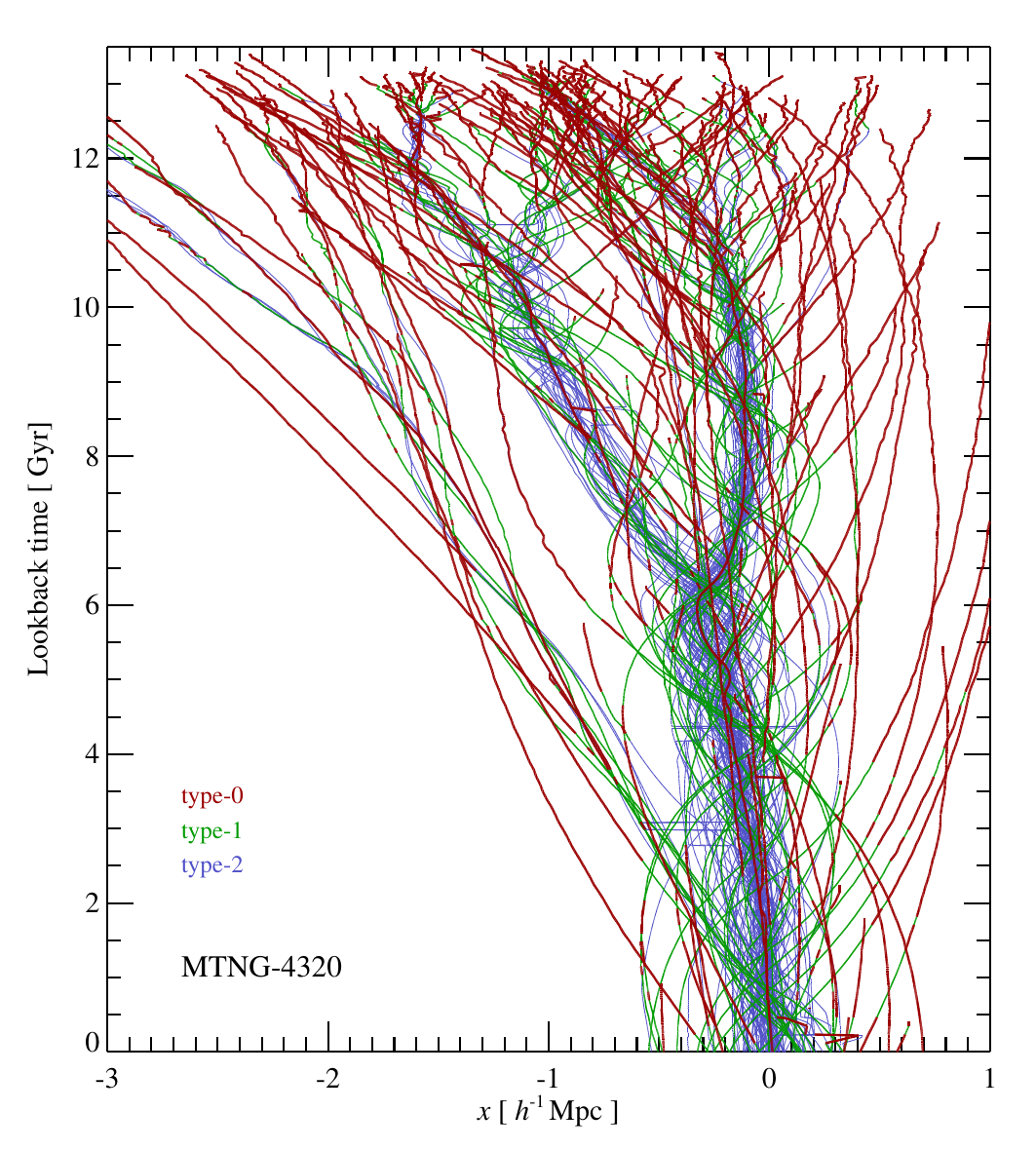}}
\caption{Example of actual galaxy orbits in our semi-analytic modelling code {\small L-GALAXIES}. We show galaxy tracks in the comoving $x$-coordinate as a function of lookback time, for a randomly picked group-sized halo of mass  $M_{200c} = 1.336\times 10^{13} \, {\rm M}_\odot$. The panel on the left is for the MTNG-2160-A simulation,  the panel on the right  for the same halo in the higher-resolution MTNG-4320-A simulation. In both cases, we distinguish central galaxies in isolated halos (`type-0') and in subhalos (`type-1'), as well as orphaned galaxies (`type-2') through the line-style, as labelled. The plot illustrates that our approach produces smooth and continuous galaxy orbits (representing the {\em actual} hierarchical merger tree). These show no obvious traces of the discreteness of the snapshot set of the underlying simulation, apart from a few rare discontinuities in some of the galaxy orbits of the higher resolution simulation. These can originate, for example, in the reassignment of a galaxy to the closest subhalo in cases where the latter lost its previously most-bound particle.
\label{fig:orbits}}
\end{figure*}

\subsection{Continuous orbits and time-integration}

In our code {\small L-GALAXIES}, all galaxies are organized as members of a certain subhalo in which they are either the central galaxy of the subhalo (then they are tracked by the most-bound particle of the subhalo, and are called `type~0' or `type~1' in the nomenclature of {\small L-GALAXIES}), or are an orphan galaxy (then their position is likewise identified by a certain particle associated with the subhalo, which was previously the central particle of a different subhalo; such galaxies are referred to as `type~2'). In either case, a particle ID is associated with the galaxy, the one that (usually) coincides with the location of the galaxy. Note that every subhalo can have only one central galaxy. If this subhalo is the most massive structure in the parent FOF group, then this galaxy is called `type~0', otherwise `type~1'.

Assume now that at some time $t_0$ for which a group catalogue is defined in the merger tree, the properties of galaxies are known. The task at hand is then to precisely define how this galaxy population is evolved to the next group catalogue's time $t_1$.  We first determine the subhalo membership of the galaxies in the new subhalo catalogue with the help of information from the merger tree.  Differently from the original version of {\small L-GALAXIES}, we do not only use descendant pointer information for this. Rather, we first determine a new provisional coordinate for a galaxy at the next snapshot time $t_1$, taken as the updated position at time $t_1$ of the particle ID that labels the galaxy.  Next, we consider a list of potential new subhalos for the galaxy, which is the union of subhalos at $t_1$ that have the galaxy's subhalo at $t_0$ as a progenitor, as well as the direct descendant subhalo of the galaxy's subhalo at time $t_0$. The galaxy is then assigned to the subhalo at time $t_1$ which has the smallest spatial distance to the provisional coordinate of the galaxy (note that this distance can also vanish if the most-bound particle of a subhalo does not change its ID, which happens quite frequently).

Next, we select which galaxy  among  the assigned ones is the central galaxy of each subhalo at time at $t_1$. If a galaxy is labelled with the same particle ID that is also the ID of the most-bound particle of the subhalo, then this galaxy is taken to be the central galaxy of the subhalo. Otherwise, the galaxy with the largest stellar mass in the subhalo that previously was a central galaxy is taken as the new central galaxy of the subhalo, or if no such galaxy is associated with the subhalo, the most massive satellite existing in the subhalo is reassigned as central galaxy in the subhalo. The new central galaxy is then changed to be labelled by the most-bound particle ID of the subhalo, which may also involve an update of its (provisional) coordinate at time $t_1$. All other galaxies in the subhalo are treated as satellite systems that are en route to merge with the central galaxy of the subhalo. Subhalos that do not contain a galaxy at the end of this assignment step get a new galaxy with zero stellar mass, no cold gas, and a hot gas mass corresponding to the universal baryon fraction assigned as central galaxy, but {\em only if} it is possible to follow this galaxy along the merger tree to $z=0$. In addition, the star formation of this new galaxy is allowed to commence only after a small random delay (a fraction of the time difference between snapshots) in order to decorrelate the creation times of new galaxies from the snapshot times.

This procedure allows galaxies to be more robustly tracked in rare edge cases, where, for example, no direct descendant has been identified or using a single descendant is unreliable because of a temporary collision of subhalos that does not yet induce an actual coalescence. Also note that unlike in older versions of {\small L-GALAXIES}, it is possible that a satellite galaxy can become again a central galaxy, while it is not possible, by construction, that a galaxy is ``lost'' (i.e.~its tracking ends before $z=0$), for example because it is created in a subhalo just above the particle resolution limit that then drops below this limit again without being linked by descendant or progenitor pointers to subsequent times. A corollary of this is that all stellar populations formed at a certain redshift are now guaranteed to be still present at $z=0$ (though they are usually reduced in mass to account for mass loss through stellar evolution).

At the end of this initial assignment step, each galaxy has a new coordinate as well as a new subhalo at time $t_1$. This allows us to define a continuous integration between the times at $t_0$ and $t_1$, which can in principle be done in a variety if ways. For the moment, we simply consider a linear interpolation of the coordinates to define the galaxy orbits. Another important aspect concerns the halo properties that are used by the semi-analytic code, which in the simplest versions is only the halo mass. For this and other subhalo properties (like the circular velocity) we also employ linear interpolation between times $t_0$ and $t_1$, in this way avoiding that the tracked subhalo properties change discontinuously in their influence on the galaxies at the times $t_0$ and $t_1$. This greatly improves the smoothness of the integration of the galaxy formation physics model, which is done by solving differential equations subject to the now time-dependent subhalo properties.  While the time derivatives of the subhalo quantities still jump at the times of group catalogue measurements, this is a second-order effect that has a much smaller influence on the results and can probably be neglected.

We also use the linear orbit integration to detect lightcone crossings of galaxies. This test is carried out in the innermost timestep loop of the physics integration of the semi-analytical model, which now also drifts the galaxies along in space, based on the linear orbit approximation. For computing the lightcone crossing itself, we use the same routines as employed in {\small GADGET-4} for ordinary particle lightcones \citep{Springel2005}. When a lightcone crossing is detected, we output the galaxy with its current physical properties. For the spatial coordinate, we have implemented two options, we either just keep the coordinate resulting from the intersection of the linear orbit approximation with the lightcone, or we replace this coordinate with a still better estimate by looking up the corresponding particle ID in the lightcone particle output of most-bound particles that we have created during the N-body simulation. Since a given particle ID can in principle occur multiple times in the latter data (due to periodicity), we use the closest occurrence the particle has to the preliminary coordinate of the lightcone crossing. Whether or not the quality of the lightcones is improved further by this additional look-up step in a significant fashion will be investigated later.

In a nutshell, the changes described above aim to decouple the time integration of the physics model (which is encoded in a set of differential equations describing, for example, radiative cooling and star formation) from the time evolution of the dark matter backbone of the structures. The latter is now realized as an updating of the halo evolution and the galaxy orbits, but without causing finite jumps at certain times.

The procedure is sufficiently involved that it can help for clarity to discuss it once more on the basis of a sketch that we show in Figure~\ref{fig:sketchsam}, illustrating key steps of our method. In this sketch we consider 5 galaxies, labelled Gal-A to Gal-E, that are distributed over two subhalos, designated as subhalo 1 and subhalo 2 at time $t_0$. Each of the galaxies is associated with a particle ID, labelled ID-a to ID-e. For the galaxies in subhalo 1, only subhalo 3 is a possible new site at time $t_1$, while for the galaxies in subhalo 2, both subhalos at time $t_1$ are possible, due to the additional progenitor link pointing from subhalo 4 to subhalo 2. This also leads to the eventual outcome here that galaxies D and E are assigned to subhalo~4, because the spatial distance of their particle IDs to the (new) particle ID tracking the center of subhalo 4, namely ID-g, is smaller than the distance to the center of subhalo 3, ID-f. Note in passing that in older versions of {\small L-GALAXIES}, the galaxies of subhalo 2 would invariably have ended up being assigned to subhalo 3 in this situation, because only the descendant pointer from subhalo 2 to subhalo 4 was considered.

Next, in subhalo 3, galaxy A is selected as central galaxy and associated with a new particle ID for tracking and for setting its position, namely ID-f. The other two galaxies stay satellites and retain their particle IDs for tracking. In subhalo 4, galaxy D is selected as central galaxy, and its coordinate is now given by particle ID-g instead of ID-d, while the other galaxy~E stays a satellite, even though it happens to be closer at that instant to the subhalo centre than the particle that used to track galaxy~D. This happens here because we first pick a new central galaxy among the ones that previously had been already a central. We also note that as part of the code's internal bookkeeping type-2 galaxies are always associated with a certain subhalo. While they can change this association in time to a subhalo other than the primary descendant subhalo, they can only `pick' the closest one in position among the subhalo set that is linked via the merger tree pointers. This in principle allows the possibility that a type-2 becomes associated with a subhalo other than the subhalo actually containing the particle, although this is exceedingly rare. After the new positions of the galaxies are determined, we obtain linear orbit approximations between the two times $t_0$ and $t_1$ (straight green lines). At their intersections with the lightcone, we obtain interpolated galaxy coordinates (red hexagons) together with galaxy properties integrated up to the corresponding times. 

In Figure~\ref{fig:orbits}, we show an example of the actual galaxy orbits obtained as a function of time when this scheme is applied to our dark matter simulations. We show the evolving galaxy population of the merger tree corresponding to a randomly selected galaxy group of virial mass $1.336 \times 10^{13}\,{\rm M}_\odot$ at $z=0$. The two panels show the formation history of the same halo at two different numerical resolutions (left is MTNG-2160-A, and right MTNG-4320-A), with tracks of galaxies in their comoving $x$-coordinate, drawn directly as they are integrated in time in our semi-analytic modelling code. Through different line colours and thicknesses, we distinguish between type-0, type-1, and type-2 galaxies, corresponding to centrals in isolated dark matter halos, centrals in dark matter subhalos, or orphaned satellite galaxies that have not yet merged with their central galaxy, respectively. The similarity in merger tree structure at the two resolutions is clearly apparent, but the higher resolution is of course able to track a much higher number of (faint) galaxies. Generically, galaxies start out as type-0 when they are a central galaxy in their own dark matter halo. When the halo becomes a dark matter substructure in a bigger structure, their track changes to a type-1 satellite system. These galaxies can sometimes merge with a (larger) galaxy, but this is usually preceded by becoming a type-2 galaxy for some time first. At several instances, we can also identify events where a type-1 galaxy becomes a type-0 again. This can happen, for example, when a substructure emerges as an isolated halo again after an interaction or fly-through with a bigger halo, yielding a `splash-back' galaxy.

\begin{figure}
    \centering
 \resizebox{8cm}{!}{\includegraphics{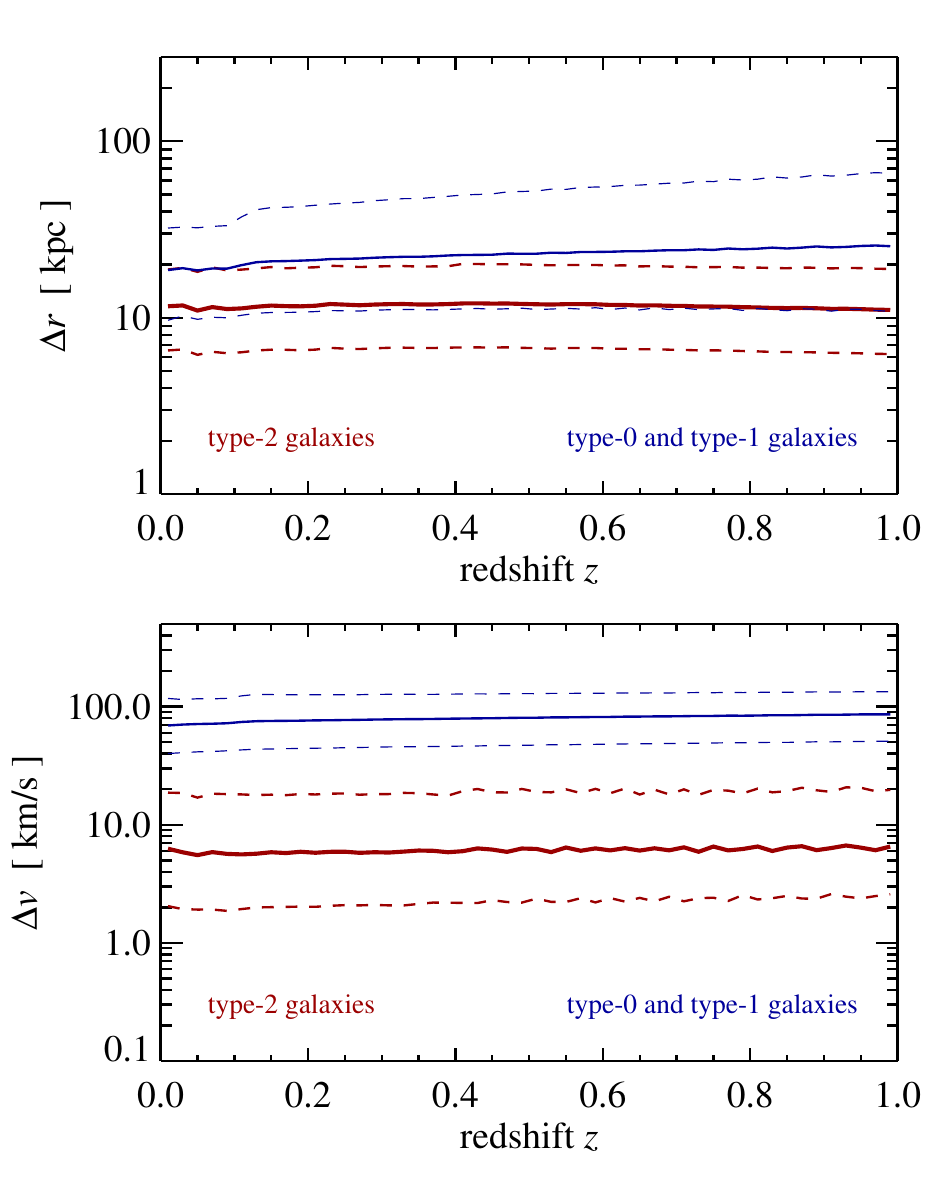}}
 \caption{Differences between lightcone crossings computed from linear orbit interpolation of each galaxy between snapshot times, and from the actual lightcone crossing of the particle identified with the galaxy at the immediately preceding snapshot time. For the given redshift range, we show the median (solid), and $16^{\rm th}$ and $84^{\rm th}$ percentiles (dashed), of the distribution of the comoving position (top panel) and peculiar velocity difference (bottom panel) in narrow redshift bins. 
   }
    \label{fig:corrections}
\end{figure}

\begin{figure*}
    \centering
    \resizebox{6cm}{!}{\includegraphics{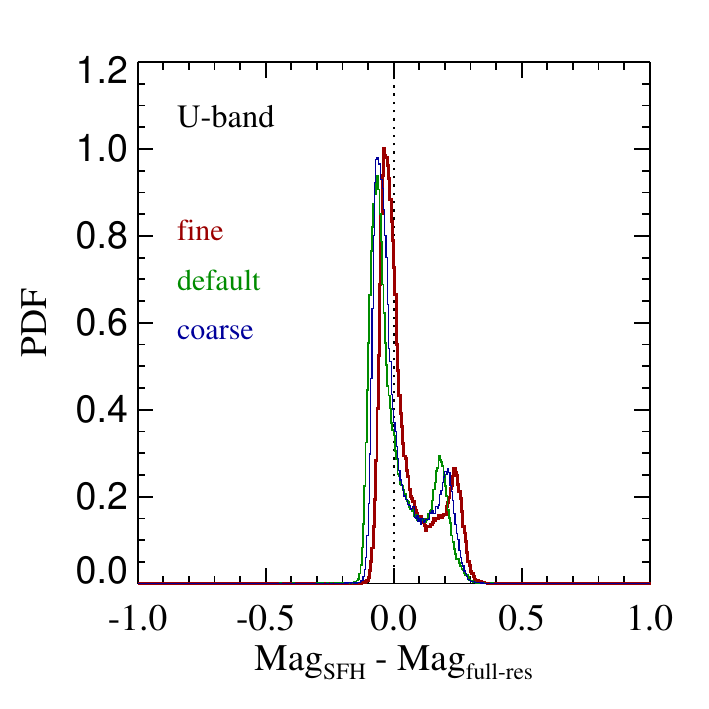}}%
    \resizebox{6cm}{!}{\includegraphics{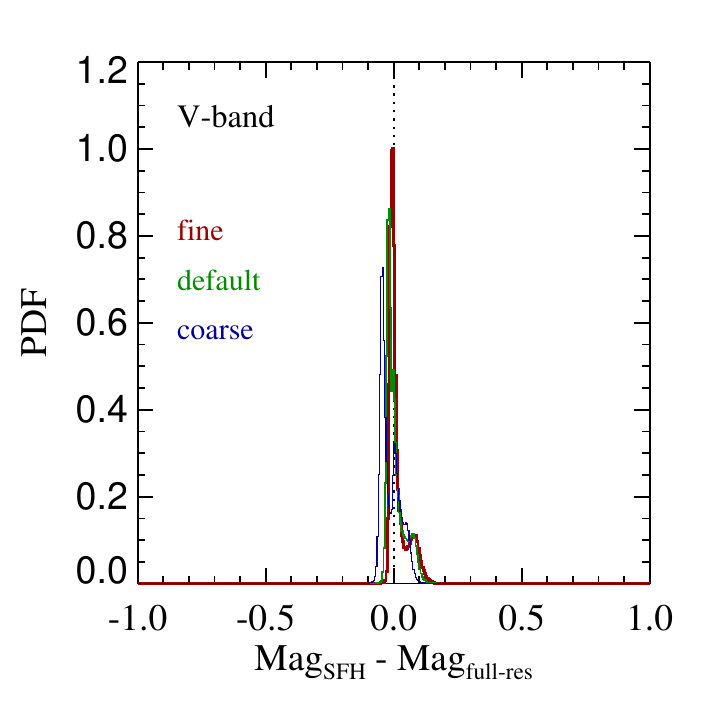}}%
    \resizebox{6cm}{!}{\includegraphics{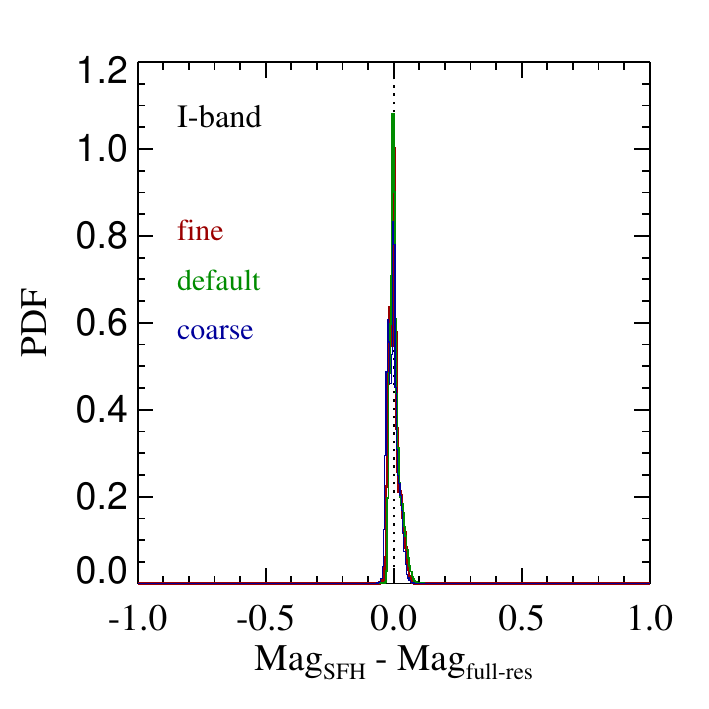}}\\
    \caption{Comparison of galaxy magnitudes in the rest frame at $z=0$, computed either from the discretized star formation history (${\rm Mag_{\,SFH}}$) or from the full time-resolution of the semi-analytic calculation (${\rm Mag_{\,full-res}}$). The different panels show results for the Johnson U-, V-, and I-band, as labelled, giving in each case the magnitude differences for the $2\times 10^5$ brightest galaxies in our $740\,{\rm Mpc}$ box. Three time resolutions are compared; `default' refers to our default choices for $t^{\rm min}_j$ and $\tau^{\rm res}_j$ (see text), while `fine' reduces these values by a factor of two, and `coarse' doubles them. Reassuringly, the outcomes are quite insensitive to these detailed choices, and the errors introduced by computing the photometry from the stored star formation history are well below $\sim0.1\,{\rm mag}$ for most galaxies, except for a few galaxies that have slightly larger errors in the U-band. } 
    \label{fig:sfh_photometry}
\end{figure*}

If desired, the spatial coordinates of the lightcone intersections can be further refined by using the particle IDs that were used to track the galaxies at time $t_0$, and then looking up their nearest lightcone crossing in the ``most-bound particle lightcone'' data produced during the N-body run. In Figure~\ref{fig:corrections}, we examine how large the corresponding corrections are for a galaxy lightcone output covering the redshift range $z=0$ to $z=1$.  We consider the difference between the lightcone crossing when the linear interpolation is used and the actual crossing obtained by looking up the particle ID used to label the galaxy's position in the stored N-body particle lightcone output.  We have subdivided the redshift range into 50 equal redshift bins, and for each redshift bin, we analyse the distribution of the differences both in comoving position (top panel) and peculiar velocity (bottom panel). Solid lines give the median for each redshift bin, while the dashed lines indicate the $16^{\rm th}$ and $84^{\rm th}$ percentiles of the corresponding distributions. We give results both for type-2 galaxies, as well as for type-0 and type-1 galaxies which still have their own dark matter subhalos.

For type-2 galaxies, the characteristic sizes of the corresponding corrections are around $\sim 11\,{\rm kpc}$ and $\sim 7\,{\rm km\, s^{-1}}$ for positions and velocities, respectively, fairly independent of redshift. These differences appear negligibly small on average, at least for the large number of outputs and thus good time resolution we have in MTNG. For the type-0 and type-1 galaxies, the values of the differences are considerably larger, and lie typically at around $\sim 20\,{\rm kpc}$ and $\sim 70\,{\rm km\, s^{-1}}$, respectively. But here the most-bound particle can be viewed as a questionable tracer anyhow, and is not necessarily expected to yield a better position and velocity for the corresponding galaxy in the first place. Recall that type 0/1s are set to the position of the minimum potential particle in a halo (this is often the same particle as the most bound one, but not always). The most bound particle ID, as well as the minimum potential ID, may also change between two output times. For the velocity, type 0/1s are set to the bulk velocity of the halo, not the velocity of a single particle. While the most-bound particle of the halo should be quite `cold' and have a small velocity relative to the halo, this velocity is not negligible and the main reason why the velocity ``corrections'' turn out to be much larger than for type-2s, simply because we here compare the bulk velocity of a whole halo with the velocity of a single particle in the halo.  We thus think that picking the position and velocity of the most-bound particle for type 0 and 1 galaxies instead of using the centres and bulk velocities of their subhalos is not expected to yield better accuracy for the lightcone crossings.

Looking up the actual lightcone crossings is thus only a worthwhile exercise for type-2 galaxies, yielding a small accuracy improvement. However, the size of this correction is so small that we consider it negligible for most practical purposes. Our default approach is therefore to work with the continuous, linearly interpolated galaxy orbits between snapshots, and to compute the lightcone crossings on the fly for these orbits. This has the additional advantage of not having to rely on a  stored N-body particle lightcone in the first place, and also eliminates the associated storage and I/O costs.

\subsection{Star formation histories and photometry}

We follow a similar strategy as \citet{Shamshiri2015} to allow magnitude reconstruction in postprocessing, but implement it in a technically different fashion. The main reason for doing this is that the snapshot spacing of the Millennium simulation project had essentially been hard-coded into the data structures used in their original implementation. We need, however, a more flexible approach, in particular because several of the simulations of the MTNG project feature a different number of outputs, as well as variable output spacing. We also want to use an overall simpler but still flexible bookkeeping scheme for the adaptive time resolution treatment, such that the storage of auxiliary information per galaxy (aside from stellar mass bins) can be avoided. Since in our new version of {\small L-GALAXIES} we process all galaxies of a tree in a strictly temporal order, it is indeed sufficient to globally specify the temporal bins used for storing the SFH of all galaxies in an identical way. Our scheme is defined as follows.

\begin{itemize}
\item 
Globally, we use an array $T^{\star,\rm end}_i$ with $i \in \{ 1, \ldots, N_{\rm bin}\}$, which defines the maximum age of stars associated with the corresponding bin $i$. The number of currently used bins is $N_{\rm bin}$ and may increase with time. Thus all stars stored in bin $i$ have an age $t_{\rm age}$ in the range  
$T^{\star,\rm end}_{i-1}  \le t_{\rm age} \le  T^{\star,\rm end}_i$,  with the implicit definition of   $T^{\star,\rm end}_{0} \equiv 0$.
\item
Each galaxy carries an individual list of stellar initial masses $M^{\star, {\rm SFH}}_i$ that encodes, together with the times defined above, the age distribution of its stellar population. Note that at later times the actual stellar mass in the bin will be lower than this because of stellar evolution  
\item
 For initialization (i.e.~at the first snapshot for $t=0$), we set $N_{\rm bin}= 1$ and $T^{\star,\rm end}_{1} = 0$.
\item
 At the beginning of every small timestep $\Delta t$ that evolves the simulated galaxy population forward in time, we increase all $T^{\star,\rm end}_i$ values by  $\Delta t$ (except for $T^{\star,\rm end}_{0}$, which is always zero). This in essence ``ages'' all already existing star formation histories of galaxies.
\item
 Stellar mass that is newly forming in a galaxy is always added to bin $i=1$ of the initial mass histogram of the corresponding galaxy.
\item 
If the maximum age of the first bin, $T^{\star,\rm end}_1$, exceeds a predefined time resolution parameter $\tau^{\rm res}$ for the SFR histories, we create a new bin. This boils down to increasing $N_{\rm bin}$ by 1, and to shifting the entries of $T^{\star,\rm end}_i$ as well as all $M^{\star, {\rm SFH}}_i$ to the element with the next higher index. As a result, $M^{\star, {\rm SFH}}_1$ becomes empty and is set to zero, while the former value of $T^{\star,\rm end}_1$ is reduced by $\tau^{\rm res}$. Note that this operation will normally happen only for a small subset of all executed timesteps.
\item
Also, at the beginning of each time step $\Delta t$ that evolves all galaxies of a tree, we check the current array $T^{\star,\rm end}_i$ for opportunities to combine two timebins into one. This is done by defining several ages $t^{\rm min}_j$ above which the corresponding time resolution of the SFR history does not have to be finer than a certain value $\tau^{\rm res}_{j}$. Our method thus checks whether for a bin $i$ there is a resolution limit $j$ with $T^{\star,\rm end}_i >  t^{{\rm min}}_j$ such that 
$T^{\star,\rm end}_{i+2} - T^{\star,\rm end}_{i} \le \tau^{\rm res}_{j}$. If this is the case, we merge the bins $i$ and $i+1$ from the SFHs by dropping $T^{\star,\rm end}_{i+1}$, coadding $M^{\star, {\rm SFH}}_{i}$ and $M^{\star, {\rm SFH}}_{i+1}$ for all galaxies, and decreasing $N_{\rm bin}$ by one, because the newly formed and enlarged bin still fulfils the prescribed temporal resolution requirements. With this approach we can flexibly guarantee any desired minimum time resolution while at the same time make the time resolution as fine as needed for younger stellar populations to still obtain accurate photometry at all times. Importantly, this operation of rebinning is done globally in the same fashion for all galaxies in a tree, i.e.~we do not have to check for this in the innermost loop over all galaxies, which is important for reasons of computational speed. 
\end{itemize}

Our default settings for defining the time resolution are $\tau^{\rm res} = 50\,{\rm Myr}$. We furthermore use 5 pairs to define the desired time resolution for older populations, as follows, $t^{\rm min}_j = \{
75$,  $150$,   $300$,  $600$, $1200\}\, {\rm Myr}$, and $\tau^{\rm res}_j = \{
50$, $100$,     $300$,    $400$, $800 \}\, {\rm Myr}$. This means that all stars formed within ages up to 75 Myr are at least represented with 50 Myr bin resolution, while for stars older than 1200 Myr, the bin resolution may drop to 800 Myr. Between these two regimes, there is a gradual transition region. With such a setting, $N_{\rm bin}$ reaches a maximum value of around 30.

In order to separately track the metallicity evolution of the stellar populations, we actually store the metal mass separately using the same bin structure. This assumes that the mean metallicity of a mass bin can be used as a good proxy to compute the photometry of the associated stellar population. This does not have to be strictly true (and every bin merger tends to reduce metallicity scatter), but our tests suggest that this approximation is sufficiently accurate for our purposes here. We also note that we store the stellar populations of the bulge and a diffuse intrahalo light component separately from the rest (which is the ``disk'' component). This thus triples the storage requirements in practice.

\begin{figure*}
    \centering
    \includegraphics[width=0.45\textwidth]{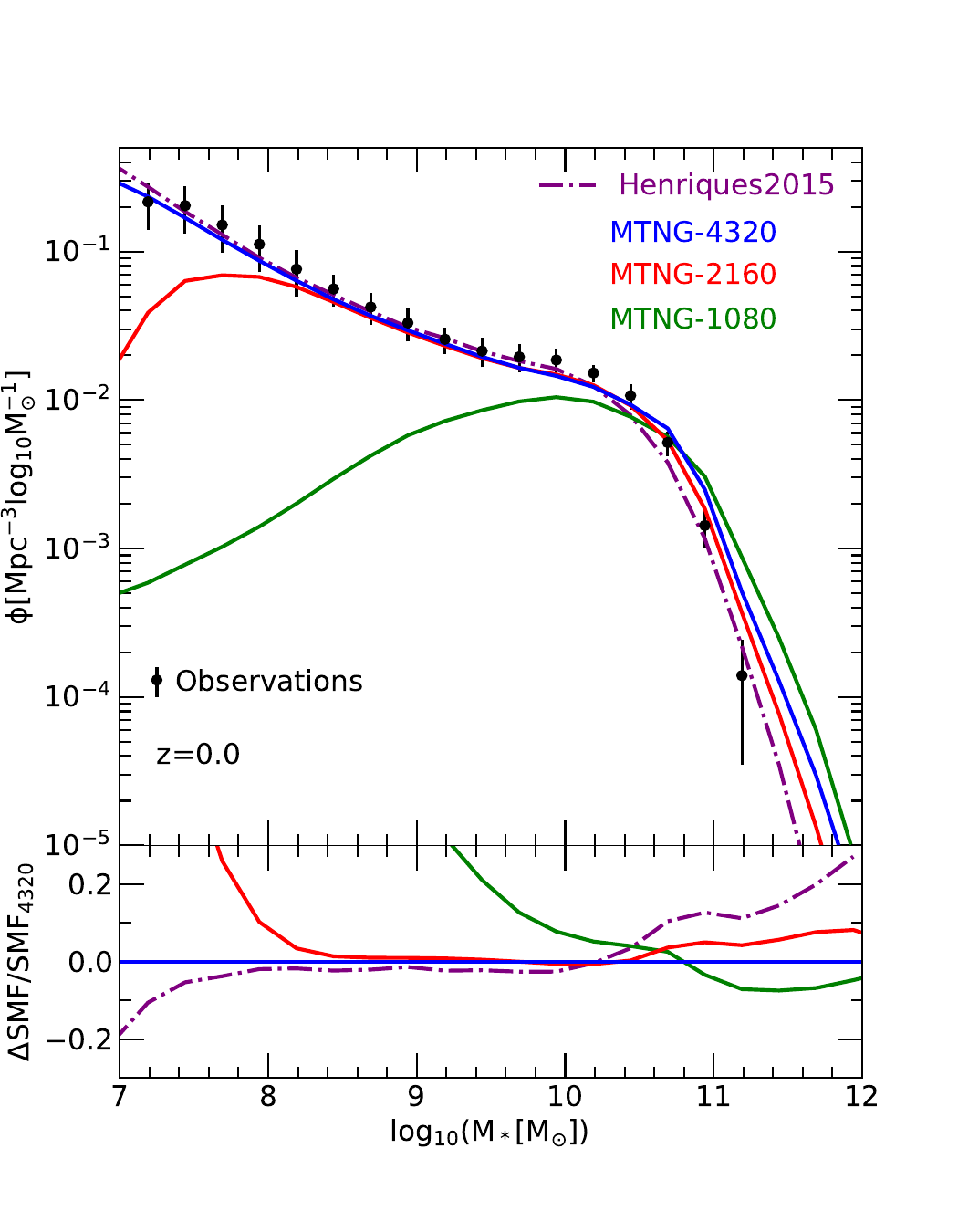}
    \includegraphics[width=0.45\textwidth]{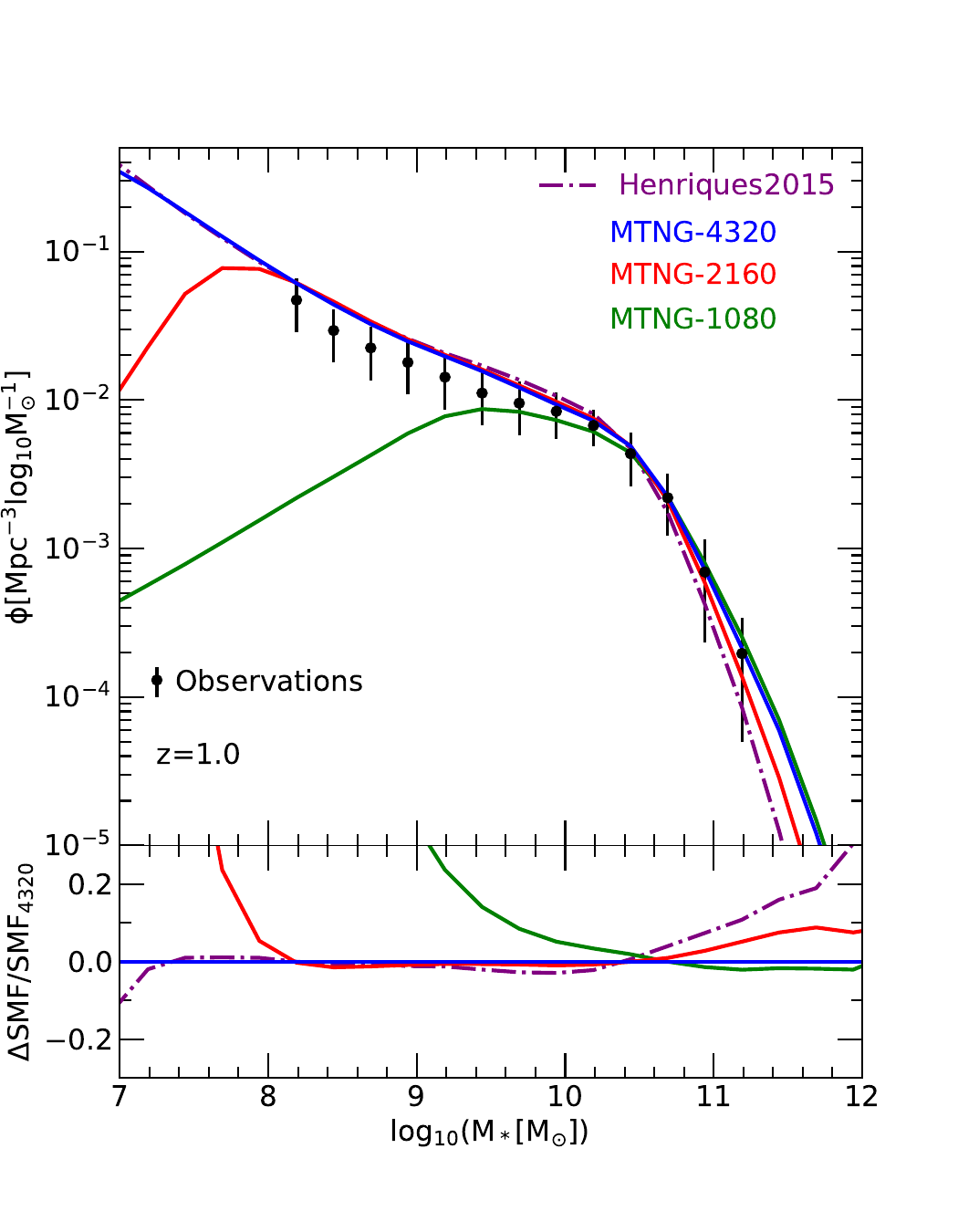}
    \\
    \includegraphics[width=0.45\textwidth]{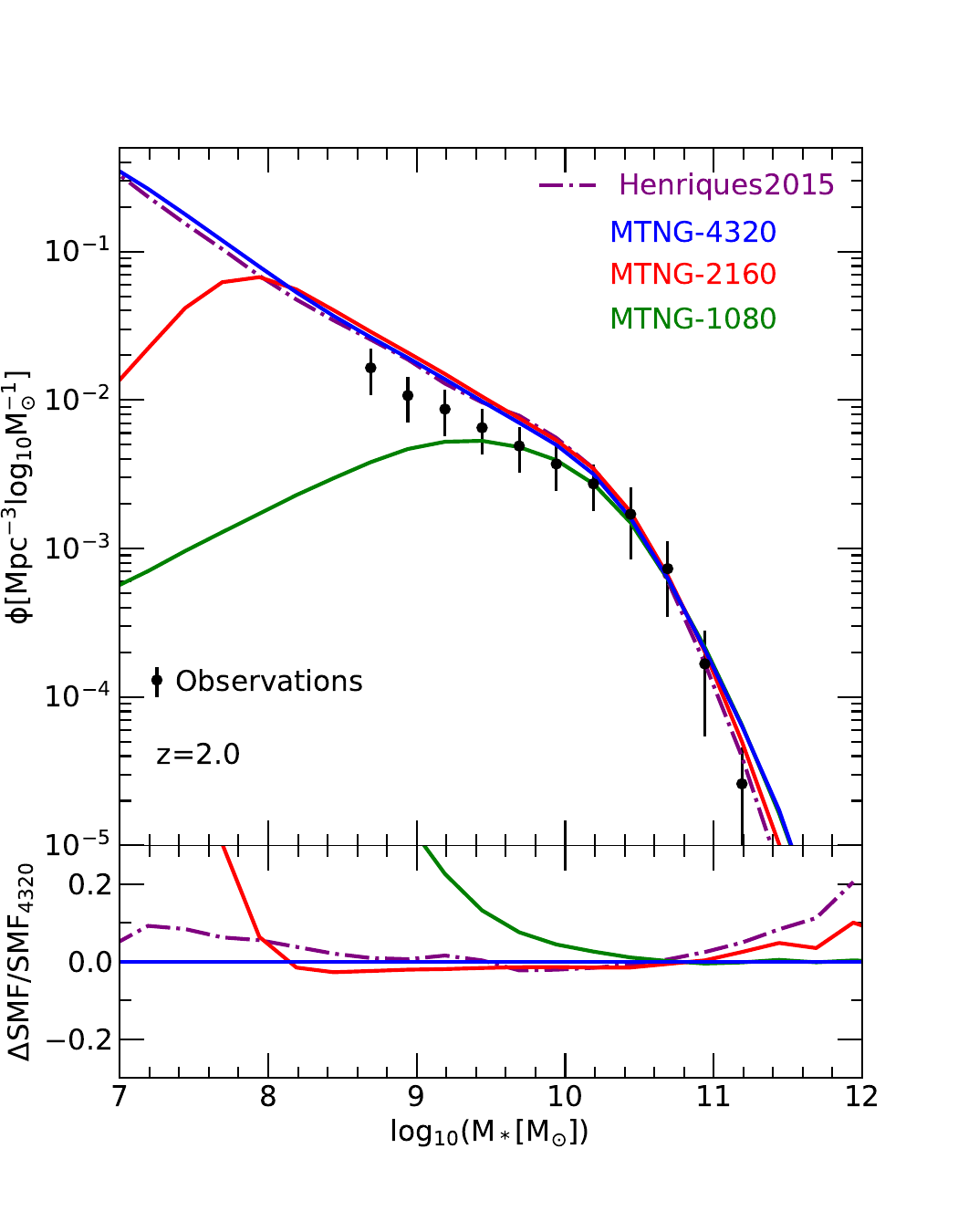}
    \includegraphics[width=0.45\textwidth]{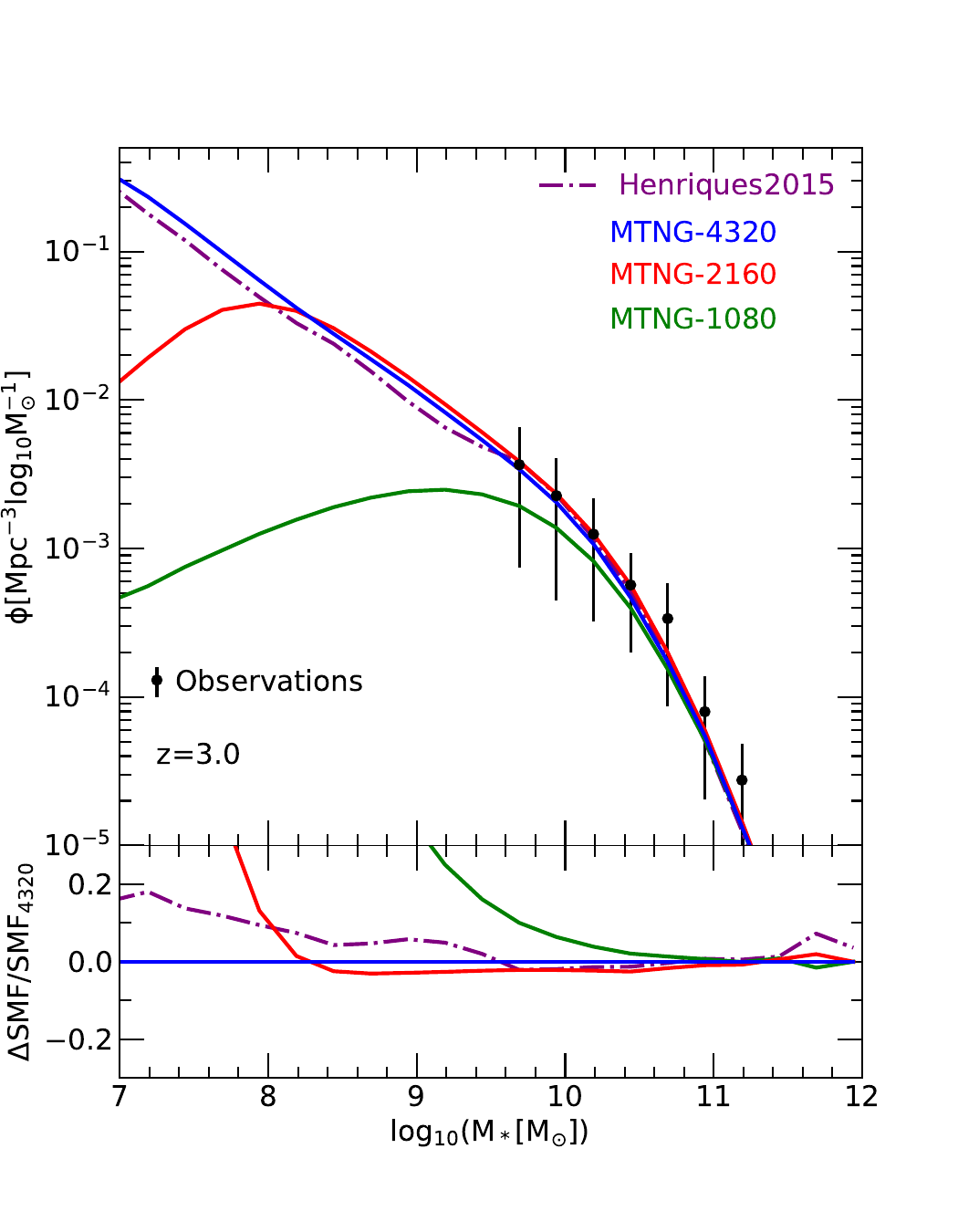}
    \caption{Stellar mass functions for three different resolutions of the MTNG740-DM simulation model, compared to the older Hen15 model and observational constraints. For each of the four displayed redshifts, $z=0$, $1$, $2$, and $3$, as labelled, the lower panel shows the difference between the stellar mass function and the result obtained with the highest resolution. Results for the A- and B-realisations are here averaged together.}
    \label{fig:smf}
\end{figure*}

\begin{figure}
    \centering
    \includegraphics[width=0.48\textwidth]{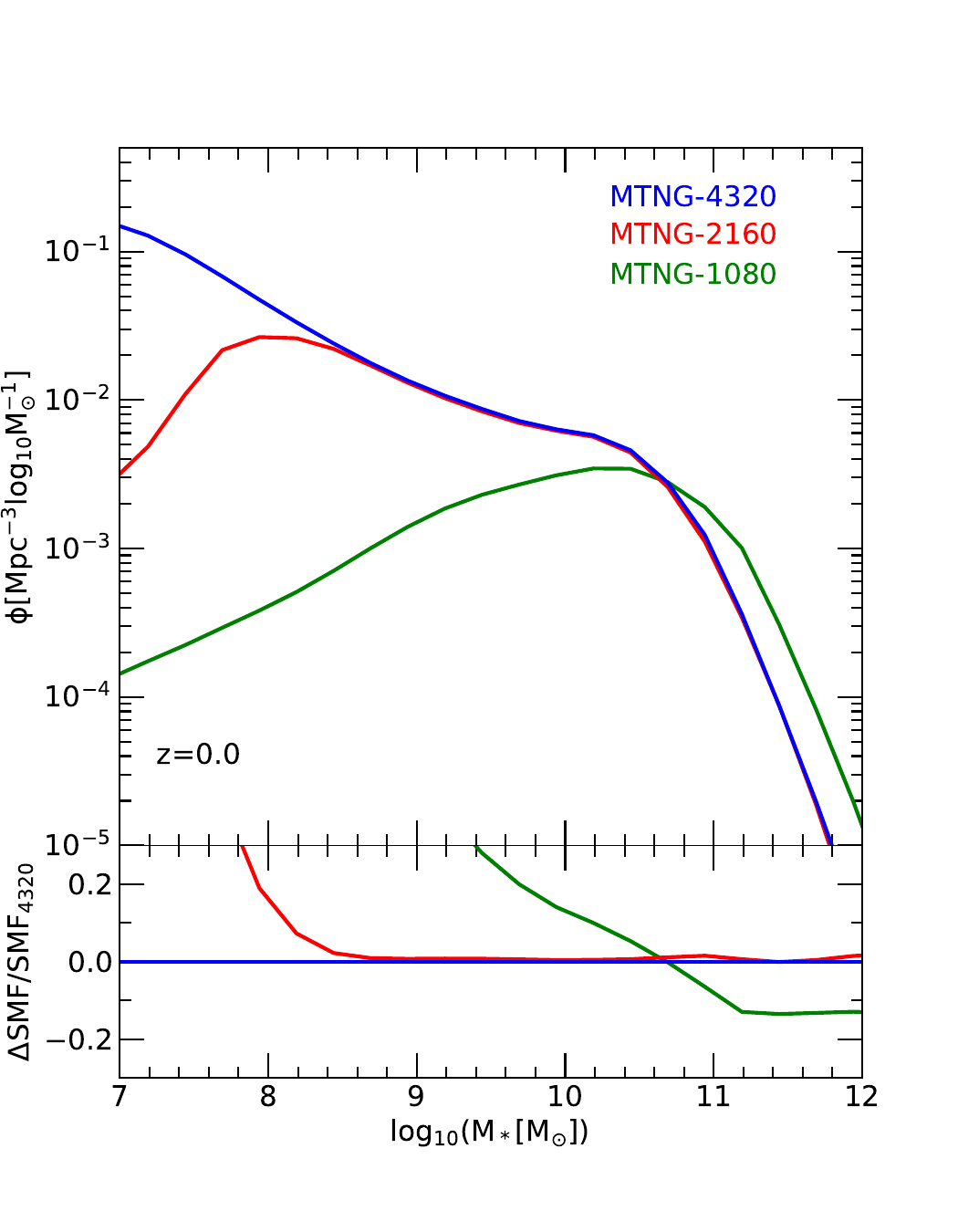}
    \caption{Stellar mass function convergence of our model when the tidal disruption treatment of type-2 galaxies is disabled. In this case, we obtain essentially perfect convergence between the 2160 and 4320 resolutions at $M_\star \gtrsim 10^8\,{\rm M}_\odot$, even at the bright end. This suggests that future refined treatments of tidal disruption should concentrate on avoiding the introduction of a residual resolution dependence due to an explicit distinction between type-2 and type-1 satellites.}
    \label{fig:smf-nodisruption}
\end{figure}

To compute the stellar luminosity in a certain observational band, we convolve the stored star formation and metallicity history of a galaxy with a stellar population synthesis model \citep[SPS, e.g.][]{Maraston2005}, 
\begin{equation}
L_{\rm band-X}^{\rm rest-frame} = \sum_{i=1}^{N_{\rm bin}} f_i \, M_i^{\star,{\rm SFH}}\,  l_{\rm band\,X}^{\rm SPS}\left(\frac{T_{i-1}^{\star,\rm end} + T_{i}^{\star,\rm end}}{2}, Z_i^\star\right)
\end{equation}
where $l_{\rm band\,X}^{\rm SPS} (t_{\rm age}, Z)$ is the luminosity predicted by the SPS in band `X' per unit initial stellar mass for a stellar population of age $t_{\rm age}$ and metallicity $Z$. The factor $f_i= ( T_{i-1}^{\star,\rm end} + T_{i}^{\star,\rm end}) / [2 (T_{i}^{\star,\rm end} - T_{i-1}^{\star,\rm end} )]  \log (T_{i}^{\star,\rm end}/T_{i-1}^{\star,\rm end})$
is an optional binning correction that largely eliminates fluctuations of the computed luminosity when age bins are merged. Since the luminosity of a stellar population is a strong function of its age, with $l_{\rm band\,X}^{\rm SPS} \propto 1/t_{\rm age}$ being a reasonable approximation for extended timespans in most bands, the young stars in any given age bin contribute considerably more than the old stars. Assuming a mean age corresponding to the bin centre for all the stellar mass of a bin therefore biases the inferred luminosity low if the formation rate has been approximately constant over the time interval. The binning correction factor removes this bias using the approximation $l_{\rm band\,X}^{\rm SPS} \propto 1/t_{\rm age}$. While this is not perfectly accurate either, it considerably reduces discreteness effects from finite bin sizes compared to simply adopting $f_i=1$. 

The result for $L_{\rm band-X}^{\rm rest-frame}$ can be directly cast to an absolute magnitude in the rest frame of the source. We consider up to 40 different filter bands. For observed luminosities, we generalise the above equation by taking the k-corrected luminosity instead, for a source at redshift $z$, computed by convolving the redshifted spectrum of the SPS with the band's transmission profile. We finally convert the k-corrected luminosity into an apparent magnitude by  including the distance modulus based on the luminosity distance to the redshift $z$ of the source.

We note that if outputting of the SFHs itself is desired, this can also be done in a storage-efficient fashion by simply outputting the current $N_{\rm bin}$ number and the $T^{\star,\rm end}_i$ values, together with the list of $M^{\star, {\rm SFH}}_{i}$ for every galaxy. Since we evolve all galaxies synchronously in time (even if located in different trees), this is possible in this fashion only for traditional time-sliced snapshot outputs. If SFHs for galaxies on a continuous lightcone output are desired, storing of  $T^{\star,\rm end}_i$ separately for every galaxy is necessary, as the binning may change whenever a new small timestep is started.

In Figure~\ref{fig:sfh_photometry}, we show a validation result for our new scheme by comparing the photometry computed based either on the discretized star formation history or doing it on-the-fly with the full time resolution of the semi-analytic computation. We give results for three different bands (for definiteness we here pick the Johnson U-, V-, and I-bands) and for three different time resolutions of the star formation history binning. Besides our default choice specified above, the `fine' case uses better time resolution by a factor of two (specified by dividing all values for $t^{\rm min}_j$ and $\tau^{\rm res}_j$ by a factor of two), while `coarse' reduces the time resolution by a factor of two compared to our default choice. Reassuringly, the scheme based on the star formation history works well overall, with typical errors  below $\sim 0.1\,{\rm mag}$, comparing quite favourably to those of \citet[][their Fig.~2]{Shamshiri2015}. As expected, the errors are largest for the U-band, due to its higher sensitivity to young stellar populations, but even here the results do not depend sensitively on the detailed choices made for the time discretization. We have checked that the errors are of very similar size if the metallicity is fixed to solar throughout; i.e.~tracking the metallicity with reduced time resolution, as required by our star formation history treatment, is not dominating the error budget.  Our default settings for the temporal resolution should thus be sufficient for essentially all applications, and there is no obvious need for further optimization. In fact, there is likely some room for a reduction in the number of required temporal bins. Conversely, for modelling nebular emission lines \citep{Hirschmann2017,Hirschmann2019}, which we do not attempt yet, a finer time-resolution for young stellar populations, e.g.~$\sim 10\,{\rm Myr}$, may still be needed, but this can be easily achieved in our formalism by changing a corresponding run-time parameter. We note that both simple \citep[see][]{Henriques2015} and sophisticated models \citep{Vijayan2019} for dust obscuration have been included in {\small L-GALAXIES}, although we do not employ them in the present work. Improving the dust modelling further and including nebular emission lines are both worthwhile areas for further work.

\section{Convergence and validation tests}
\label{sec:convergence}

For this study, we kept the physical parameters of the semi-analytic model at the values
determined by \citet{Henriques2015}, who had used the stellar mass function and the red fraction of galaxies at four different redshifts as constraints to set the parameters. This allows us to assess whether or not the extensive changes and upgrades we implemented in our methodology have a significant impact on the results. Furthermore, we are here primarily interested in examining the numerical convergence of the semi-analytic model, and in particular, to establish which mass resolution is required to reach accurate results down to a prescribed stellar mass limit. Further improving the physical modelling will be addressed in forthcoming work.

In Figure~\ref{fig:smf}, we show results for the stellar mass function at four different redshifts, $z=0$, 1, 2, and 3, obtained with our new version of {\small L-GALAXIES} applied to the MTNG740-DM simulations. We show in each case averaged results for the A- and B-realisations, but restrict ourselves to the $1080^3$, $2160^3$, and $4320^3$ resolutions, as still lower resolutions turn out to be inadequate even at the bright end. We compare both to the old \citet{Henriques2015} results and to the observational constraints used by them.

Reassuringly, we find generally quite good agreement between our new results and the older ones based on combining the Millennium and Millennium-II simulations. This is despite the extensive changes of the underlying numerical methods, which involved everything from the N-body simulation code, the group finding and merger tree construction algorithms, to the semi-analytic code itself. This speaks for the general robustness of the approach, and can be viewed as an important validation of the new methods themselves.
 
In terms of convergence with mass resolution, for the $1080^3$ run (our `level-3' resolution, which has a dark matter particle mass of $m_{\rm dm} =1.26\times 10^{10}\,{\rm M}_\odot$) we find acceptable results only for stellar masses at the knee of  the stellar mass function and higher, $M_\star \ge 10^{10}\,{\rm M}_\odot$, while the faint end is basically completely missing. For  the $2160^3$ resolution (`level-2' with $m_{\rm dm}=1.57\times 10^{9}\,{\rm M}_\odot$), we achieve numerical convergence to substantially fainter limits, $M_\star \ge 10^{8}\,{\rm M}_\odot$. This will already be sufficient for most practical applications to galaxy surveys, which usually target substantially brighter galaxies. In contrast, for hydrodynamical simulations of galaxy formation such as IllustrisTNG, this resolution would still be too low to produce meaningfully accurate results. Finally, for our $4320^3$ model (`level-1', $m_{\rm dm}=1.96\times 10^{8}\,{\rm M}_\odot$), the accuracy down to the faintest galaxies is excellent, and we conservatively estimate that the galaxy abundance for  $M_\star \ge 10^{7}\,{\rm M}_\odot$ is fully converged. At the bright end, we find residual small convergence problems for the $2160^3$ and $4320^3$ runs at low redshift, $z=1$ and $z=0$. We have found that these are largely due to the treatment
of tidal disruption of satellites in \citet{Henriques2015}, which was originally introduced in \citet{Guo2011}. As this modelling is only applied to type-2 galaxies it carries an implicit dependence on numerical resolution because more galaxies can be followed as type-1 satellites when the resolution improves, and furthermore, it is applied on a discrete basis at snapshot times, giving it an implicit dependence on the spacing of outputs that makes it  difficult to  mesh with our new continuous time integration approach. If we disable this physical model, we obtain perfect convergence also at the bright end, as we show explicitly in Figure~\ref{fig:smf-nodisruption}. This suggests that it will be worthwhile to develop an improved disruption model as part of future studies, for example, following the gradual stripping scenario proposed by \citet{Henriques2010}.

\begin{figure}
    \centering
    \includegraphics[width=0.48\textwidth]{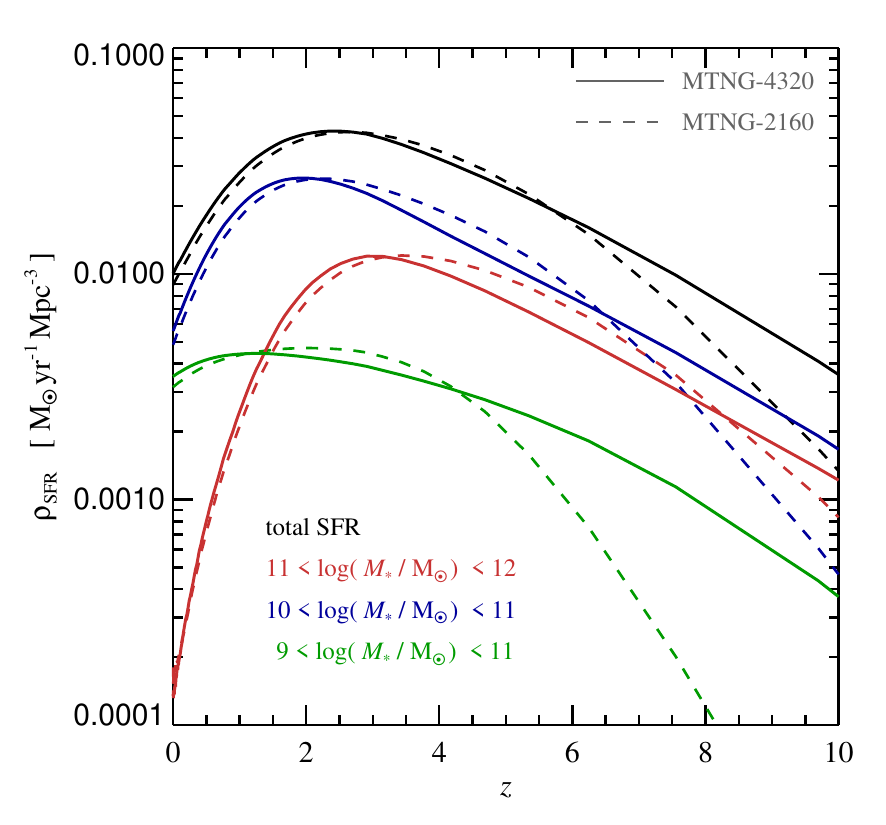}
    \caption{Cosmic star formation history for the full galaxy population and for samples of galaxies of different stellar mass selected at $z=0$, as labelled. We compare results for the MTNG-4320 (solid lines) and MTNG-2160 (dashed lines) resolutions.  While the convergence is good at late times, where most of the cosmic time lies and thus most of the stellar mass forms, the MTNG-2160 model is unable to resolve the small mass halos that dominate star formation at very high redshift.}
    \label{fig:sfh}
\end{figure}

At the bright end of the stellar mass functions in Figure~\ref{fig:smf}, there are also noticeable differences between MTNG and the result of \citet{Henriques2015} for the Millennium simulation. While this is likewise reduced if the disruption treatment of type-2 galaxies is disabled, the difference here is not unexpected as it can already arise from the substantial difference in cosmology between the two models, in particular in the baryon fraction of halos, which was $\Omega_{\rm b}/\Omega_0 = 0.045/0.25 = 0.18$ for the Millennium project, whereas it is $\Omega_{\rm b}/\Omega_0 = 0.0486/0.3089 = 0.1573$ for the Planck cosmology adopted in MTNG.

A view of the temporal build-up of the stellar mass is given in Figure~\ref{fig:sfh}, where we show the cosmic star formation rate density as a function of redshift, both for the total galaxy population, as well as for galaxies in three different mass bins selected today at $z=0$. What is shown for these latter samples is the actual star formation history of the corresponding galaxies (including also `ex-situ' stars that merged into the galaxies). This type of analysis is possible thanks to the stored star formation histories of each of our semi-analytic galaxies. We compare results for the MTNG-4320 and MTNG-2160 resolutions, so the plot also serves as a further convergence test. Reassuringly, the convergence is generally quite good, both for the total star formation rate density as well as for the star formation histories of the galaxy samples of fixed stellar mass today, at least this is true for low redshift where most stars form. However, at high redshift, $z\gtrsim 5$, the star formation rate in the low-resolution model is suppressed compared to the higher resolution simulation. At these early times the star formation density is dominated by low-mass halos that are not properly resolved in the MTNG-2160 simulation, so this is to be expected. With time, the halo mass scale that dominates star formation shifts to larger halos \citep{Springel2003}, allowing MTNG-2160 to eventually catch up and yield converged results for the bulk of the galaxies at late times. Another well-known result evident from the plot is that more massive galaxies have older stellar populations, and that their star-formation has shut-off earlier than that of low-mass galaxies. This seemingly anti-hierarchical behaviour contrasts with the hierarchical growth of the dark matter halos themselves \citep[e.g.][]{DeLucia2006}.

\begin{figure}
    \centering
    \includegraphics[width=0.48\textwidth]{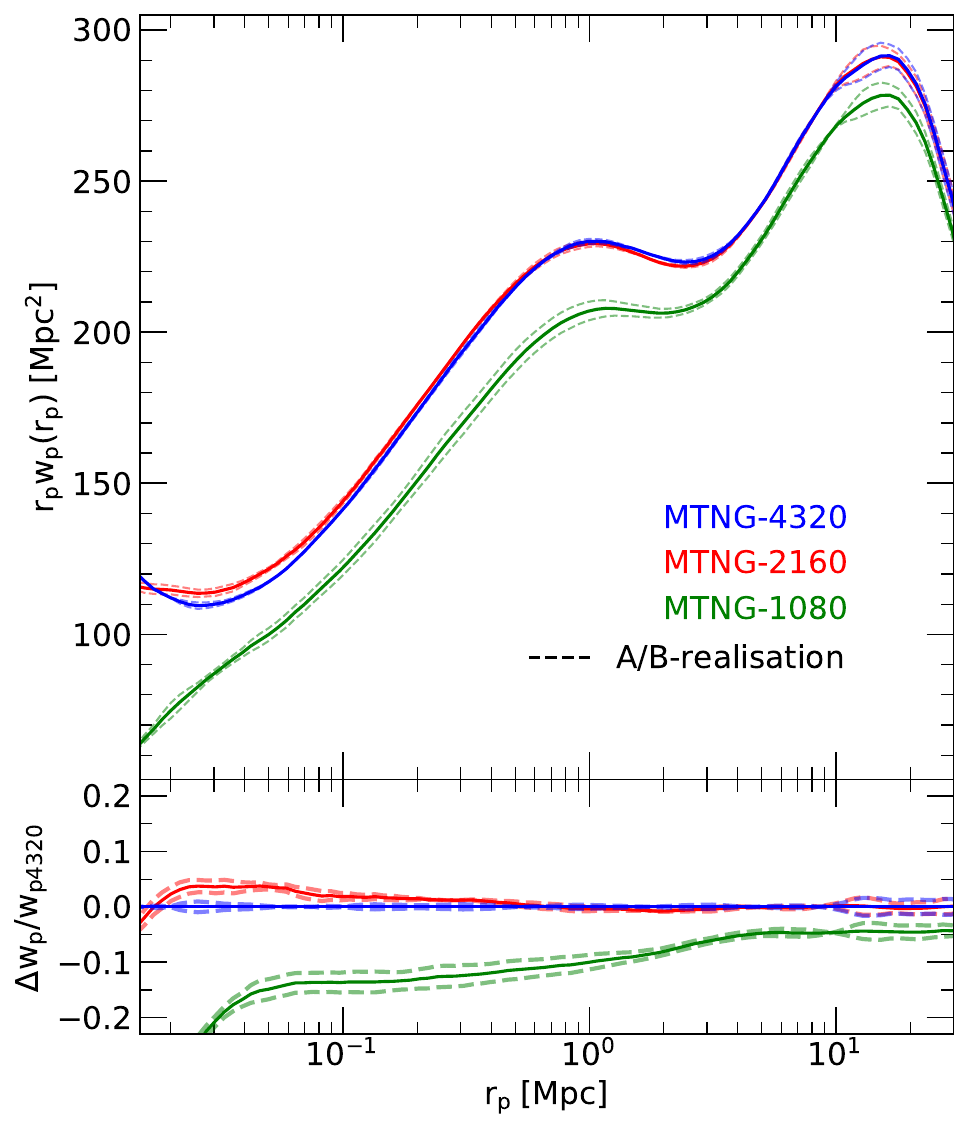}
    \caption{Projected correlation functions at $z = 0$ for galaxies with stellar mass greater than $10^{10}\,{\rm M}_\odot$ at three different numerical resolutions of the MTNG740-DM model}, as labelled. The lower panel shows the difference between each resolution and the highest one. The solid lines were obtained from averaging the results for the A- and B-realisations in each case. The dashed lines correspond to the individual A and B runs.
    \label{fig:clustering_snaps}
\end{figure}

\begin{figure}
    \centering
    \includegraphics[width=0.48\textwidth]{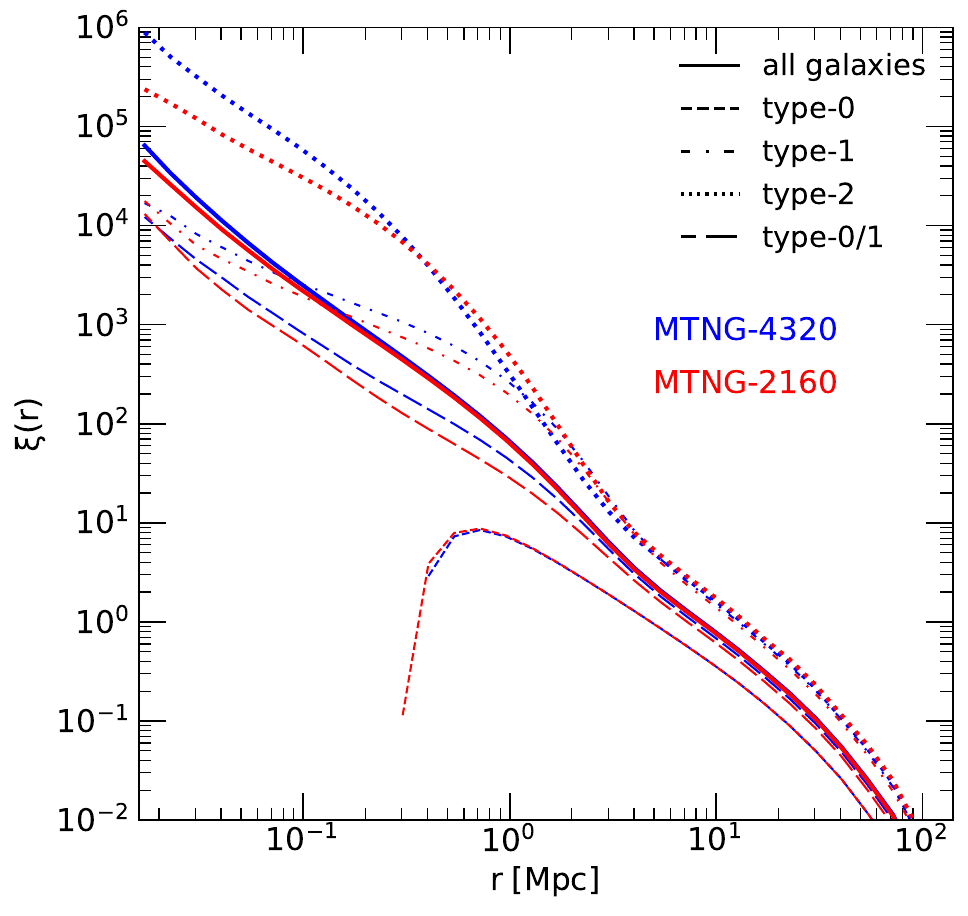}
    \caption{Real-space two-point correlation functions for galaxies with $M_* > 10^9\,{\rm M}_\odot$ are compared for the MTNG-4320 and MTNG-2160 simulations. We show results for the full galaxy sample (solid lines), as well as separately for type-0 (dashed), type-1 (dot-dashed), and type-2 (dotted) galaxies. Type-0 galaxies are the central objects of resolved  FOF groups and thus show a small-scale exclusion effect. This is no longer visible in the results for the combined sample of type-0 and type-1 galaxies, which traces all resolved  dark matter subhalos.  Note that type-2 orphan galaxies must also be included to obtain a converged small-scale clustering signal between MTNG-4320 and MTNG-2160.}
    \label{fig:clustering_types}
\end{figure}

\begin{figure*}
    \centering
\resizebox{17cm}{!}{\includegraphics{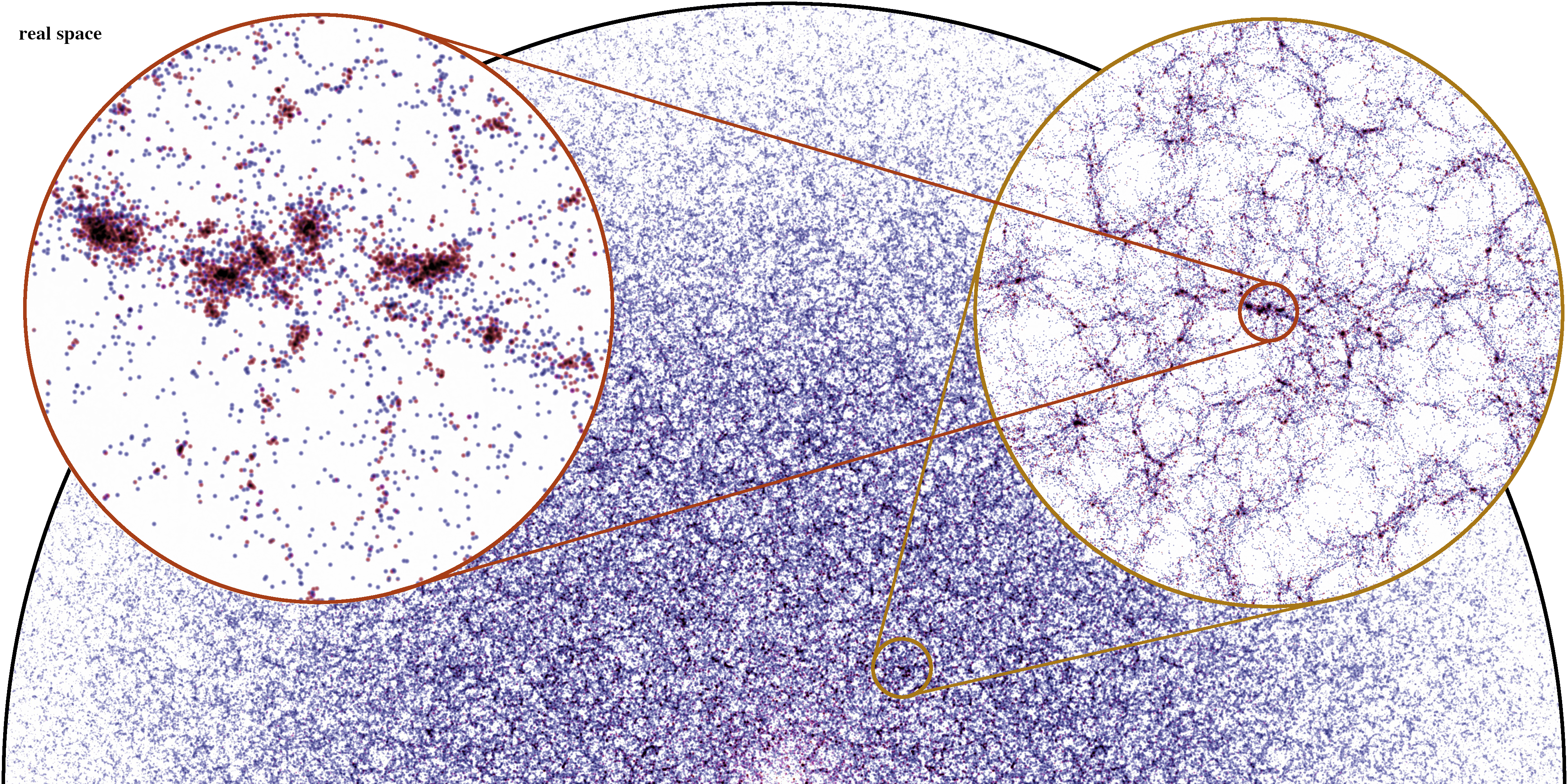}}\vspace*{-0.05cm}\\%
\hspace*{-0.04cm}\resizebox{17.4cm}{!}{\includegraphics{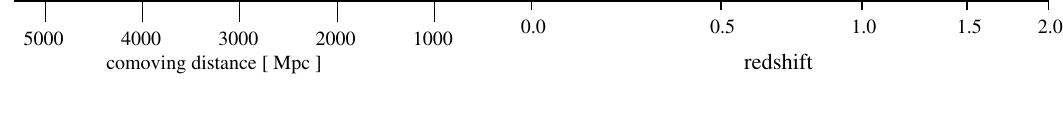}}\vspace*{-0.7cm}\\%
    \caption{Galaxy distribution on the past lightcone  of the MTNG740-DM-1A simulation} down to Johnson apparent magnitude $R < 23$, in a 180~degrees wide, thin wedge with opening angle 0.24 degrees, out to redshift $z=2$. The galaxy positions are drawn as circles with comoving coordinates in real space, using a red color hue for galaxies with rest frame color index $B-R > 0.7$, and a blue color hue otherwise. The two circular insets show nested zooms with diameters of 400~Mpc and 40~Mpc, and fainter apparent magnitude limits of $R < 25$ and $R < 28$, respectively. In these insets, the projection thickness is constant at 21.14~Mpc (slightly thicker than the background image), matching exactly the geometry of corresponding images of the dark matter distribution in \citet[][their Fig.~1]{Aguayo2022}.
    \label{fig:180degreesA}
\end{figure*}

\begin{figure*}
    \centering
\resizebox{17cm}{!}{\includegraphics{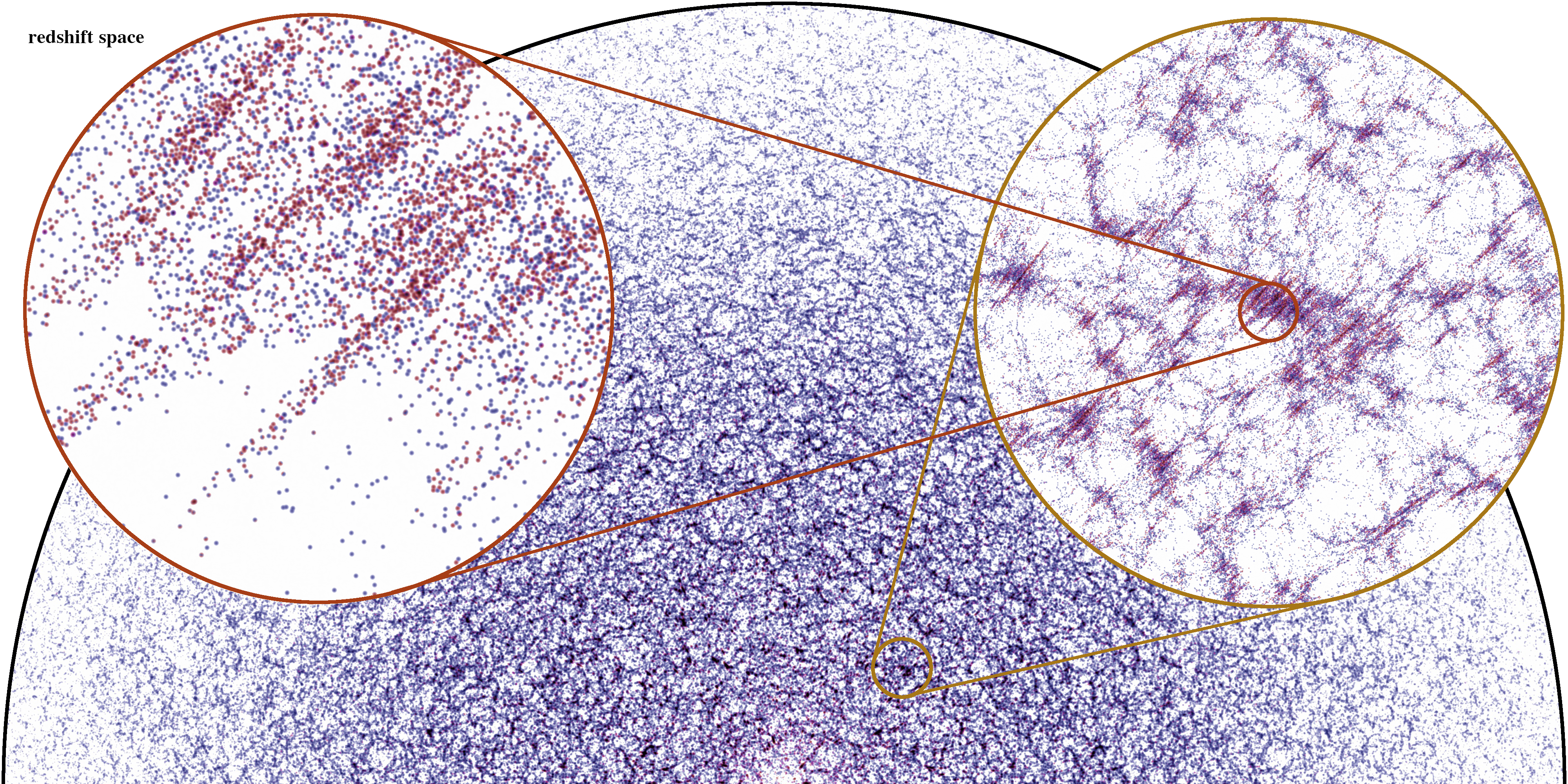}}\vspace*{-0.05cm}\\%
\hspace*{-0.04cm}\resizebox{17.4cm}{!}{\includegraphics{images/pie_axis.pdf}}\vspace*{-0.7cm}\\%
    \caption{Like Fig.~\ref{fig:180degreesA}, except that the galaxy positions are drawn in {\em redshift space}. A visual comparison of the clustering patterns in the two figures readily makes the effects of redshift space distortions apparent, producing a squashing of structures along the line-of-sight direction on large scales due to infall, and a stretching on small scales (``fingers-of-god'') due to random motions in virialized objects.}
    \label{fig:180degreesB}
\end{figure*}

In Figure~\ref{fig:clustering_snaps} we consider the numerical convergence of clustering predictions at $z=0$, measured in terms of the projected two-point correlation function in redshift space (in Section~\ref{sec:clustering} we describe the procedure used to calculate these projected correlation functions).  We include all galaxies with stellar mass above $\mathrm{10^{10}\,M_{\odot}}$. Comparing to our highest resolution result, we find convergence to better than the 1-2 percent level for  $r_{\rm p} \ge 400\,{\rm kpc}$ at $2160^3$ `level-2' resolution. On smaller scales, clustering in the level-2 case gets progressively stronger than in level-1, although the difference is still only $5$ percent by  the remarkably small scale of $20\,{\rm kpc}$. Since this is deep in the one-halo regime of clustering, it reflects the fact that the distribution of satellites around host galaxies reproduces extremely well, being only slightly more concentrated in the lower resolution simulation. In contrast, the low resolution level-3 run shows a systematic underprediction of the clustering strength which is about 4-5 percent in the two-halo regime beyond  $3\,{\rm Mpc}$, growing to nearly 10 percent at edge of the one-halo regime at $r_{\rm p} \sim 700\,{\rm kpc}$ before yet larger differences appear at smaller scale. This suggests an underprediction of the total number of satellites, together with further increased concentration of the satellite distribution. The relatively small offsets of the dashed lines from the solid lines in the bottom panel show that all these systematic trends are substantially larger than `cosmic variance' scatter between realisations. Nevertheless, it is interesting to see that on larger scales (in the two-halo regime) averaging the A and B realisations clearly produces a substantially smoother result than expected just from the improvement in statistics, thus demonstrating the value of using paired simulations. 

It is also interesting to examine the contributions of the different types of galaxies to the clustering signal. In Figure~\ref{fig:clustering_types} we show the real-space two-point correlation functions of all galaxies with stellar mass $M_\star \ge 10^9\,{\rm M}_\odot$ at $z=0$. We compare results for the MTNG-4320 and MTNG-2160 simulations and show in each case not only the result for the correlation function of all galaxies (sold lines), but also separately for type-0, type-1, and type-2 galaxies, and, in addition, for the combined sample of type-0 and type-1 galaxies. The correlation function for the full galaxy sample is close to a power-law and converges nearly perfectly, as is also the case for the type-0 galaxies. As there can only be one central type-0 galaxy per FOF group, the clustering signal of type-0's shows a significant short-range exclusion effect. Evidently, clustering predictions on small-scales, in the one-halo regime, require proper accounting for satellite galaxies. The type-1 and type-2 galaxies exhibit clear signatures of the one- and two-halo regimes, with type-2's showing the strongest small-scale clustering and the highest large-scale bias overall, due to their preferential presence in the most massive dark matter halos. Interestingly, the type-1 and type-2 clustering signals do not converge individually between MTNG-4320 and MTNG-2160, only their combination does. As a consequence, the correlations for galaxies associated with  gravitationally bound subhalos (i.e.~type-0 and type-1 combined) also do not converge as well as the total galaxy sample including also type-2's. This underlines the importance of including orphan galaxies for small-scale clustering predictions \citep[see also][]{Guo2011, Guo2014}.

\section{Galaxies on the lightcones}
\label{sec:gallightcone}

In this section, we consider the galaxy lightcone output made possible by the new version of {\small L-GALAXIES}. We will first give a visual impression of the continuous lightcone output, which is easily amenable to the imposition of an apparent magnitude limit. By additionally considering galaxies in redshift space, the realism of predicted galaxy mock catalogues can be greatly increased. We shall then analyse the projected galaxy clustering for lightcone galaxies, and compare to measurements obtained from ordinary time-slices (i.e.~snaphots). In particular, we check whether it makes a difference for the results whether snapshots or the continuous lightcone output is used.

\subsection{Pie diagrams}

In Figure~\ref{fig:180degreesA}, we show galaxies selected down to apparent magnitude $R < 23$, in a thin wedge that is 180 degrees wide and has an opening angle of just 0.24 degrees. The galaxies are depicted at their real comoving distance out to redshift $z=2$. Galaxies with rest frame colour index $B-R > 0.7$ are largely quenched and are shown with a red circular symbol, while all other galaxies are drawn with a blue symbol. The background image gives a nice visual impression of the cosmic web traced by the galaxies. The contrast of this web becomes noticeably weaker towards higher redshift. This is not just a result of the decline of the tracer density with redshift, but also reflects the fact that the structures are not as pronounced and non-linear at higher redshift. This can also be readily appreciated by comparison to the corresponding dark matter distribution in our companion paper by \citet[][their Fig.~1]{Aguayo2022}.

We also show two insets in Fig.~\ref{fig:180degreesA} that successively zoom in to a rich supercluster region and are 40~Mpc and 400~Mpc in diameter, respectively. The intermediate enlargement gives a clear illustration of the filamentary large-scale structure, with the biggest concentrations of galaxies found at their intersection points. It is apparent that red galaxies are preferentially found in these group- and cluster-sized concentrations, an impression that becomes particularly evident in the final enlargement.

In Figure~\ref{fig:180degreesB}, we display the galaxies in the same viewing geometry but in redshift space. The background image now shows a subtle difference in the clustering pattern in the form of a squashing of structures along the line-of-sight, caused by coherent infall onto large-scale structures. This is one aspect of the well-known redshift space distortions. The other becomes prominently visible in the intermediate scale zoom, where there is a pronounced stretching of virialized structures along the line-of-sight due to the associated internal random motions, the so-called `finger-of-god' effect. The final zoom makes it clear that,  as a result of this effect, correctly identifying membership in bound structures is substantially more difficult in redshift space than in real space.

\subsection{Clustering on the lightcone}
\label{sec:clustering}

\begin{figure*}
    \centering
    \includegraphics[width=0.95\textwidth]{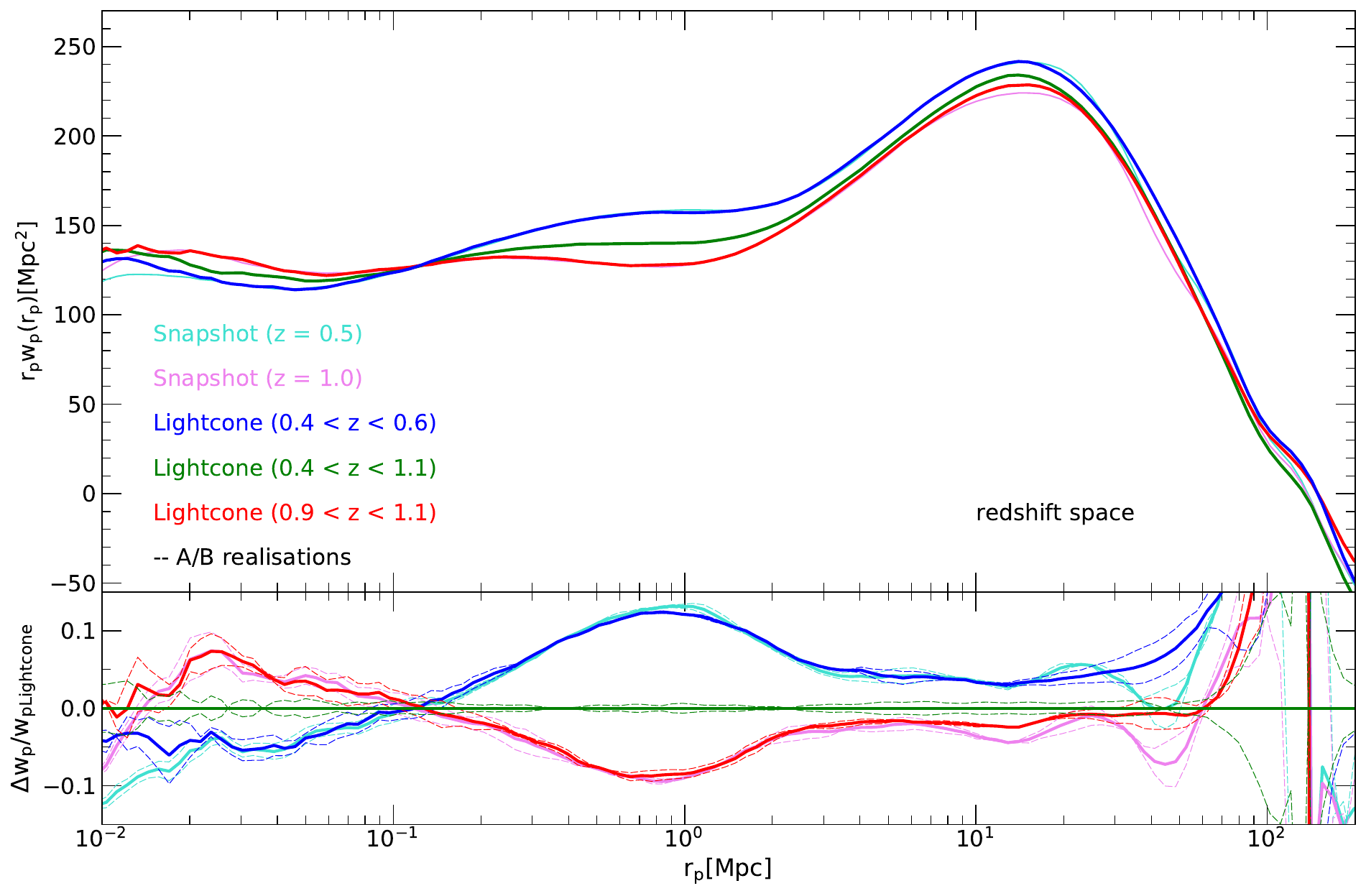}
    \caption{Projected correlation functions for galaxies with stellar mass above $\mathrm{10^{10}\,M_{\odot}}$ in our MTNG740-DM-1 simulations}, both on the full-sky past lightcone, and for snapshots at fixed time. For the lightcones we show results for galaxies in three different thick  redshift shells. In the lower panel, we show the difference between each measurement and the one obtained for the thickest lightcone shell. In all cases, the solid lines were obtained by averaging results for the A- and B-realisations, while the dashed lines show these individually.
The results for the snapshots can be compared with those obtained for the thinner lightcone shells. The fact that these do not agree precisely illustrates the small but systematic differences which arise when lightcone samples are approximated using single snapshots. Likewise, the differences between the three lightcone shells highlight the influence of evolution of galaxy clustering with redshift. 
    \label{fig:twoPoint}
\end{figure*}

\begin{figure}
    \centering
    \includegraphics[width=0.49\textwidth]{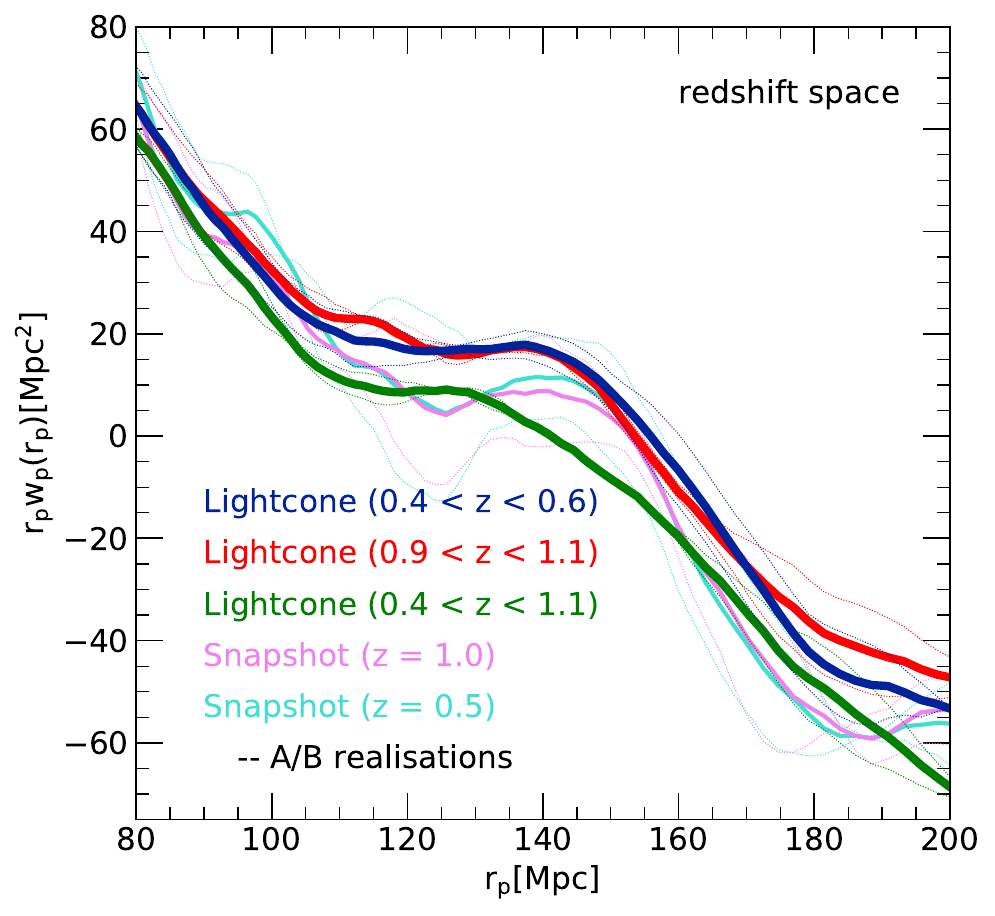}
    \caption{Large-scale clustering around the scale of the baryonic acoustic oscillations (BAO), as seen in the projected two-point correlation functions obtained from our MTNG740-DM simulations} using galaxies both in shells of the full-sky past lightcone and in two snapshots at fixed redshift. One of the shells extends over $0.4< z < 0.6$, a second over $0.9< z < 1.1$, and the last over $0.4 < z < 1.1$. In all cases, the solid lines were obtained by averaging the results for the A- and B-realisations, while the dotted lines show the two realisations separately. The systematic difference between the thinner lightcone shells and the snapshots on which they are centred is a result of the anisotropy of the large-scale  autocorrelation function induced by the cubic periodic geometry of the simulation. The differences between the lightcone shells themselves are driven by residual cosmic variance, which is present despite their significant comoving volume, as evidenced in more detail in Appendix~\ref{appendix:lightconeclustering}. We note that the binning here has been chosen somewhat finer than our default logarithmic bin size, in order to yield smoother curves over the relative narrow radial range shown.
    \label{fig:twoPoint_zoom}
\end{figure}

An important practical issue for modelling the expected clustering signal of galaxies is the question of whether measurements based on timeslices at certain fixed redshifts give sufficient accuracy (for example, taking the snapshot at the centre of the redshift range of an observational sample), or whether one has to use proper lightcone output to get a precise enough result. In Figure~\ref{fig:twoPoint}, we address this question by comparing estimates of the projected two-point galaxy correlation function, $w_{\rm p}(r_{\rm p})$ for redshift shells within our all-sky lightcones to similar estimates made using snapshots centred on the narrower shells. The projected correlation function has the advantage of being comparatively insensitive to redshift-space distortions. Also, it can be directly measured for observational data, and it is only a function of one variable, the transverse separation $r_{\rm p}$ of galaxy pairs.

To calculate the projected correlation function in redshift space we first add the contribution of the line-of-sight (LOS) peculiar velocity to the comoving distance of a galaxy. For snapshots, each of the three principal coordinate axes is chosen  in turn as the LOS, while for lightcones we take it to be the radial direction along which the galaxy is viewed. For a chosen target galaxy, we project the 3D redshift-space separation vector to each neighbouring galaxy onto the target's LOS vector, thus defining the parallel separation $\pi$ of the two galaxies. Then $r_{\rm p} \equiv  (r^2 - \pi^2)^{1/2}$ is taken to be the corresponding transverse separation. The redshift-space correlation function $\xi(r_{\rm p}, \pi)$ can then be measured by the natural estimator
\begin{equation}
\xi(r_{\rm p},\pi) = \frac{\rm D_t D}{\rm R_t R} - 1,
\end{equation}
where ${\rm D_tD}$ is a symbolic short-hand for the mean number of companion galaxies in a small volume element ${\rm d}r_{\rm p}{\rm d}\pi$ around $(r_{\rm p},\pi)$ for a randomly selected member of the target galaxy sample ${\rm D_t}$, while ${\rm R_tR}$ stands for the mean number of companion galaxies in the same volume element relative to the target galaxy if the positions of companion galaxies are randomized while preserving their mean spatial density. For the projected two-point correlation function, we then calculate
\begin{equation}
w_{\rm p} (r_{\rm p}) = \int_{-\pi_{\rm max}}^{\pi_{\rm max}} \xi(r_{\rm p},\pi)\, {\rm d}\pi , 
\end{equation}
where we pick $\pi_{\rm max} = 370\,{\rm Mpc}$, i.e.~half the size of our MTNG740 simulation box. With this choice {\it all} distinct projected pairs are counted in the snapshots. Using logarithmic bins in radius, $w_{\rm p} (r_{\rm p})$ can be directly estimated for each bin centre $r_{\rm p}$ by averaging over all target galaxies the number of companions in a hollow cylindrical volume with inner bin radius $r_{\rm p}^-$ and  outer bin radius $r_{\rm p}^+$ and total length of $2\, \pi_{\rm max}=740\,{\rm Mpc}$ parallel to our LOS.

In the case of snapshots, the target set ${\rm D_t}$ is identical to D, and the mean count of random background galaxies simply follows analytically from the number of galaxies in the periodic simulation box. To further reduce the measurement uncertainty, we separately determine  correlation function estimates for projections along the $x$-, $y$-, or $z$-axes, averaging the three results. For our lightcone measurements, the set D consists of all galaxies in a full-sky lightcone over the redshift range $0.2 < z < 1.25$, while for the target galaxies we choose a substantially narrower redshift slice, for example $0.4 < z < 0.6$. This avoids any edge effects since the cylindrical volume over which companions are counted never overlaps a redshift boundary.  Unlike in the snapshot case, it is not here possible to estimate the comparison random counts analytically, because the  galaxy population evolves with redshift, and hence the mean number density of galaxies which pass our selection criterion (here $M_{\star} > 10^{10}\,{\rm M}_\odot$) is also a function of redshift. We address this problem by creating a random sample out of the actual lightcone data by randomising the angular positions of all the galaxies while keeping their comoving distances fixed. This produces an effectively unclustered sample while retaining the radial variation of mean galaxy density. 

Note that with this definition of the lightcone estimator we aim to reproduce that used on individual snapshots as closely as possible, while also eliminating edge effects of the kind  identified by \citet{Nock2010}. If one were to create a lightcone by replicating a single snapshot periodically through all space, our lightcone estimate would be identical with that from the snapshot itself, except on scales where the simulation autocorrelation function becomes anisotropic because of the cubic periodicity of the simulation. We will see below that such anisotropic effects are easily detectable at the BAO scale in our MTNG740 simulations because  projection directions are isotropically distributed in our full-sky lightcones but are always parallel to the principal directions in our snapshots. 

For measuring the average counts of neighbouring galaxies we have typically employed 150 logarithmic bins in $r_{\rm p}$ between 1.5~kpc and 370~Mpc (except for the close-up analysis of the BAO region in Fig.~\ref{fig:twoPoint_zoom}, where we used still finer bins). For every chosen target galaxy, we determine the exact number count with the help of a parallel tree-based algorithm that hierarchically groups the galaxies. If a node of the tree falls fully inside a bin, all galaxies of the node can be counted at once and the tree walk along the corresponding branch can be ended. This allows an efficient computation of the correlation function at large distances even if the size of the set D is very large. For the measurements presented here, we have however, for computational convenience, downsampled the number of primary targets to 2 million if ${\rm D_t}$ was bigger than this number.

In Figure~\ref{fig:twoPoint}, we first compare clustering  in a lightcone shell with $0.4 < z < 0.6$ to that in a snapshot at the middle\footnote{For the $z=0.4$ to 0.6 redshift interval, the mean comoving-volume weighted redshift is $z=0.5141$, which we here approximate with the mean arithmetic average of the shells' boundary redshifts, and similarly for the  $0.9 < z < 1.1$ interval.} of this interval, $z=0.5$. The  former has substantially lower noise than the latter because of its substantially larger volume. Interestingly, the clustering measured in these different ways agrees very well for distances $r_{\rm p}$ between 20 kpc to 20 Mpc (to $\sim 1$ percent accuracy or even better). The situation degrades slightly at separations below 20~kpc, where differences of up to 10 percent show up. We have verified that these can be attributed to small inaccuracies in the orbit interpolation of our semi-analytic code, which affect the positions of galaxies on the lightcone. Using fewer snapshots than we have employed increases these differences, consistent with this explanation. 

For $r_{\rm p}$  in the range 20 Mpc to 100 Mpc,  larger deviations of size 5-10 percent show up. In this regime also the differences between the A- and B-measurements become noticeably larger, indicating that here the results become sensitive to cosmic variance, to the precise way in which the averaging and projection of the correlation function are done, and to exactly which part of the snapshot volume is mapped into the lightcone shell. Finally, the relative differences between lightcone-shell and snapshot results become quite large at distances beyond 100~Mpc, chiefly because the correlation function itself becomes very small, anisotropic and eventually even negative there. We will examine this region separately below.

The story is very similar if we compare in Figure~\ref{fig:twoPoint} the redshift shell $0.9 < z < 1.1$ to a single snapshot measurement at $z=1.0$. Again, the results are in excellent agreement for $r_{\rm p} < 10\,{\rm Mpc}$, but the relative difference grows to several percent at larger separations, and becomes large for $r_{\rm p} > 100\,{\rm Mpc}$, although the shape of $w_{\rm p}(r_{\rm p})$ is well tracked even there. We note in passing that using galaxy samples selected by apparent magnitude instead of absolute stellar mass will likely introduce additional subtle differences between lightcone shells and snapshots at fixed redshift, due to the k-corrections involved. These effects are expected to depend on the chosen band and the colour of the galaxy sample. We thus defer their analysis to forthcoming work where we intend to construct mock catalogues that closely match the observational characteristics of upcoming galaxy redshift surveys.

\begin{figure*}
    \centering
    \includegraphics[width=0.95\textwidth]{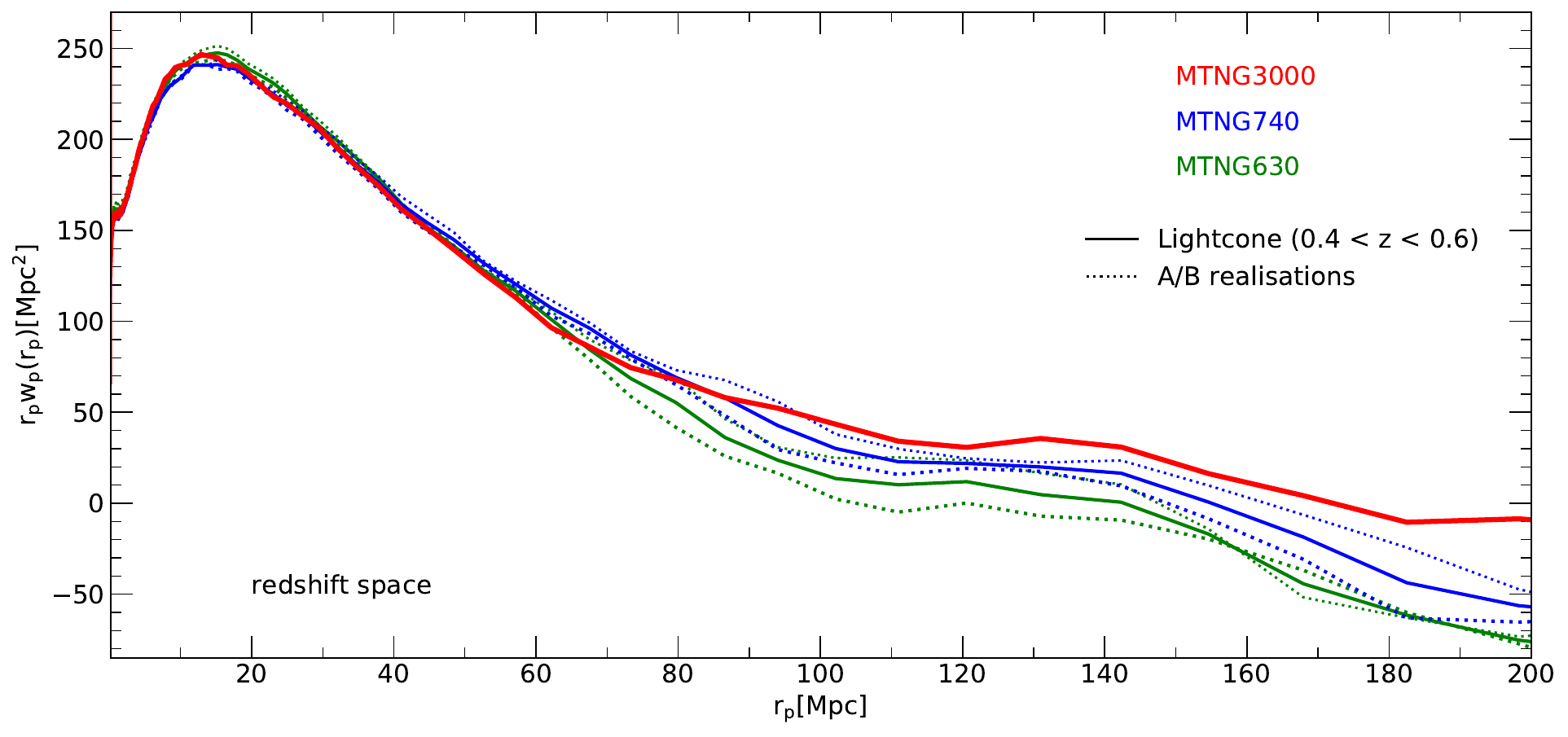}
    \caption{Projected correlation function of galaxies in the redshift range $0.4 < z < 0.6$ and with stellar mass above $\mathrm{10^{10}\,M_{\odot}}$, calculated from the full-sky lightcones of simulations with different box sizes. The model MTNG3000 refers to our simulation with a box size of $3000\,{\rm Mpc}$ and $10240^3$ dark matter  as well as $2160^3$ massive neutrino particles. Its cosmology is  slightly different from our default MTNG740 model, which is shown for comparison. We also include our MTNG630 simulation, which has the same cosmology and mass resolution as MTNG3000, except for a smaller box size of 630~Mpc, thus reliably indicating effects due to boxsize alone. We show the average signal of the A and B realisations as solid lines for the two small boxes, and the dotted lines show the results for their A/B realisations individually. For the big box, we have at this point only the A realisation, which is shown as a solid line. }
    \label{fig:xxl}
\end{figure*}

Lightcone shells  over different redshift ranges in general yield different clustering signals. Figure~\ref{fig:twoPoint} emphasises this by also including a clustering estimate for a thick redshift shell, $0.4 < z < 1.1$, which can be compared with the two narrower shells discussed above, i.e.~$0.4 < z < 0.6$ and $0.9 < z < 1.1$. There are substantial differences in shape between the three estimates. The largest differences occur at $r_{\rm p}\simeq 1\,{\rm Mpc}$, and are more than 10 percent between the lowest and highest redshift shells. Also, at the smallest and largest scales shown, there are differences exceeding $5$ percent. Not unexpectedly, the results for the thick shell lie between those for the low- and high-redshift shells near its edges. These results thus stress  that correlation functions change shape significantly over the redshift ranges spanned by real large galaxy surveys . These changes are due both to changes in the statistics of the mass distribution itself, and to changes in how galaxies occupy dark matter (sub)halos. Both aspects must be treated accurately and consistently across the full redshift range if robust predictions are to be made at the percent level. This will be possible with snapshots only if many are stored and they are appropriately and consistently combined.
This is automatically achieved using lightcone outputs for semi-analytic models as we do here.

For the lightcone results in the lower panel of Figure~\ref{fig:twoPoint} it is noticeable that both the bin-to-bin noise and the difference between the A and B realisations are much smaller at separations below a few Mpc than the systematic shape difference between $z\sim 0.5$ and $z\sim 1.0$. At larger scales the bin-to-bin noise remains small but a systematic offset appears between the two realisations. In Figure~\ref{fig:twoPoint_zoom} we show how these results extend to even larger scales around the BAO feature and the zero-crossing of the projected correlation function. Here we plot both axes linearly and again include results for the two realisations separately. There is no sign of significant bin-to-bin ``noise'' either in the individual realisations or in their means, and the shape of the five measurements is quite similar. There are, however, substantial differences in  amplitude between the different realisations. Interestingly, the overall amplitude of the BAO feature at $\sim$140~Mpc  is quite similar for the two snapshot results, and also for the two thin redshift shells, but these two pairs are significantly offset from each other. This is a consequence of the anisotropy of the galaxy autocorrelation function on these scales which is a significant fraction of the periodic scale of the MTNG740 simulations. As a result, $w_{\rm p}(r_{\rm p})$ is systematically different when the LOS is parallel to one of the principal axes (as for the snapshots) than when it is averaged over all possible directions (as for the lightcones). As we show explicitly in Appendix~\ref{appendix:lightconeclustering}, the reason why the thick redshift slice yields a result that is systematically different from the thin slices can be attributed to cosmic variance effects, because the coverage of the lightcone shell by replicas of the periodic simulation box is not uniform in the sense that not all points internal to the simulation box are mapped an equal number of times into the lightcone shell. Averaging over an ensemble of lightcone origins (i.e.~different observer locations in the box) would be able to eliminate this effect (see Appendix \ref{appendix:lightconeclustering})

It is clear that the finite box size of MTNG740 influences  correlation function estimates on large scales, both because of missing long wavelength modes and because of the anisotropic distribution of the long wavelength modes which are included. Furthermore, the correlation function within a periodic box must obey an integral constraint: its integral over the whole simulation volume must be identically zero. This forces the estimate of $w_{\rm p}(r_{\rm p})$ to cross zero at a separation which is typically 20 to 30\% of the box-size, even if the correlation function correponding to the theoretical initial linear power spectrum crosses zero at significantly larger scale. If the box size is too small, this integral constraint can result in a substantial underestimate of the true zero-crossing scale. Figure~\ref{fig:xxl} demonstrates this effect for our default box size; we compare correlation function estimates for the lightcone shell $0.4 < z < 0.6$ for our flagship MTNG3000 simulation, which has a $3000\,{\rm Mpc}$ box size, to estimates for our MTNG630 simulations which have identical cosmological and simulation parameters, except that the box size is just $630\,{\rm Mpc}$. While the two estimates agree well on scales below about $70\,{\rm Mpc}$, the smaller box gives systematically smaller correlation estimates on larger scales, crossing zero at about $140\,{\rm Mpc}$ rather than at the correct value  of $\sim 170\,{\rm Mpc}$ as found for the large box. 

Note that both these simulations include massive neutrinos with a summed mass of 100meV, while the simulations analysed elsewhere in this paper  have zero neutrino masses and a slightly different cosmology. These differences also impact structure on large scales, as we demonstrate by including in  Figure~\ref{fig:xxl} the estimate of $w_{\rm p}(r_{\rm p})$ for the MTNG740 simulations which we discussed previously. This lies significantly above the estimate for MTNG630 on scales exceeding  $50\,{\rm Mpc}$. As expected, the inclusion of non-zero neutrino mass significantly affects the predicted shape of the galaxy correlation functions. While all three models in this plot show a BAO feature at the same spatial scale, there are systematic differences in correlation amplitude on these large scales.  In forthcoming work, we will study the impact of neutrinos on the clustering signal in substantially more detail, and will also complement the  MTNG3000 run shown here with a corresponding B-realisation.

Taken together, Figs.~\ref{fig:twoPoint}, \ref{fig:twoPoint_zoom} and \ref{fig:xxl} show that while averaging our two realisations allows us to substantially reduce random fluctuations in our clustering measurements for individual snapshots, it is effective only on intermediate scales (up to few tens of Mpc) in our all-sky lightcones. For the latter, the large effective volume, corresponding to averaging over many line-of-sight directions through the simulation box, substantially reduces fluctuations. However, this cannot prevent significant systematic effects due to the finite box size and limited total number of long-wavelength modes. These must be understood and analysed carefully when interpreting real surveys.

\section{Summary and Conclusions}
\label{sec:conclusions}

In this paper, we introduced a major modification of the {\small L-GALAXIES} semi-analytic code for galaxy formation, making it suitable for application to the new simulations of the MillenniumTNG project, and capable of producing smooth lightcone output that is largely free of discreteness effects from the underlying finite set of group catalogues. We have also described how the improved merger tree structure of MTNG can be suitably exploited to enhance robustness of the tracking of galaxies.  

The central element of our new approach lies in a better time integration of the semi-analytic physics models used by the code. In particular, we have eliminated the possibility of discontinuous steps at snapshot times in essentially all the quantities relevant for galaxy evolution, including the positions and velocities of galaxies (thus their orbits), and quantities such as the virial radii of halos (which are relevant for cooling and feedback). In previous versions of the code, such jumps could occur whenever a new group catalogue was fed into the model. This was not problematic as far as the accuracy of final quantities goes, provided outputs were only generated for the snapshot times themselves. A continuous outputting strategy, however, as needed for the lightcones, can make the discontinuities visible. This compelled us to work on eliminating them.

Our solution for computing the lightcone-crossings of galaxies is based on finding the intersection of linearly interpolated galaxy orbits and the lightcone. If desired, the resulting phase-space coordinates can be further refined by looking up a stored lightcone crossing (as output by the underlying N-body simulation) of the most-bound particle used to track the galaxy's position. We have found, however, that the corresponding corrections are very small in practice for the high output time resolution we have in MTNG, making this step optional. Eliminating it offers additional flexibility, in that an N-body particle lightcone is no longer needed for the semi-analytic processing. {\small L-GALAXIES} can then produce a lightcone on the fly with its origin at any desired position, using the same techniques as the {\small GADGET-4} N-body code. In particular, the simulation box will be periodically replicated if needed to fill the prescribed lightcone geometry, and multiple different lightcones with different geometries and redshift ranges can be created at the same time.

We have shown that our new code produces semi-analytic predictions that are already converged at much worse mass resolution than would be needed for a full hydrodynamical simulation. This is a major advantage, as it not only saves a huge amount of CPU time, but also allows the semi-analytic code to be applied to moderate resolution N-body simulations covering extremely large volumes, Gpc box-sizes and beyond. For example, in the MTNG project, we have completed a calculation with a $(3000\,{\rm Mpc})^3$ volume and more than 1.1 trillion particles. Its mass resolution is nearly twice as good as our $2160^3$ run of the MTNG740-DM model. We expect this simulation to be ideal for applying our new semi-analytic methodology, since it will allow galaxy catalogues to be made over the full sky and to high redshift down to stellar masses below $10^8\,\mathrm{M}_\odot$, yet including a very large number of Fourier modes even beyond the BAO scale. We will address this task in forthcoming work.

Our initial analysis of the projected two-point clustering of galaxies shows clearly that clustering in real galaxy samples cannot be modelled to high accuracy using sparsely spaced snapshots at fixed times. Predictions that are accurate at the percent level can only be achieved with properly constructed lightcones including a consistent and sufficiently flexible galaxy formation model. Fortunately, the advantages of the ``fixed-and-paired'' technique carry over to measurements of galaxy clustering signals on lightcone shells, reducing the cosmic variance that would otherwise show up more prominently in simulations with moderate box size, such as those we have shown here. Nevertheless, significantly larger boxes will be needed to achieve fully accurate results on the BAO scale and beyond. Another point we have emphasised is that redshift boundaries on the lightcone can significantly bias the shape of clustering measurements, especially at large separations such as the BAO scale. Proper interpretation of observations therefore requires careful forward modelling of the data, taking selection and boundary effects accurately into account.

In comparison to their predecessor \citet{Henriques2015}, our new mock catalogues are based on simulations carried out in a cosmology with parameters in good agreement with recent estimates; the underlying merger trees were constructed with an improved algorithm that better tracks substructure, resulting in a more robust galaxy evolution model; furthermore, we stored about four times as many snapshot catalogues as for the Millennium Simulations, greatly improving the time resolution of the trees. We are able to make galaxy catalogues on the full-sky past lightcone  with galaxy evolution and clustering treated smoothly, continuously and in a physically realistic fashion over a large redshift range (see  Figure \ref{fig:sfh}). This also allows us to calibrate our models using observations from current and future large surveys. The much larger effective volumes they now provide will also  allow us to carry out more precise large-scale structure analyses than was previously possible (e.g. Figure \ref{fig:xxl}).

In this paper, we have deliberately avoided updating the physical assumptions of the SAM described in \citet{Henriques2015}\footnote{Aside from addressing a few minor weaknesses such as the treatment of galaxy disruptions.}, a task that we intend to tackle in forthcoming work that should also address important issues such as realistic dust modelling. As part of such improvements, we  intend to quantify the uncertainties in SAM predictions, in particular, those relevant to galaxy clustering in wide-field surveys, due to uncertainties in physical parameters or modelling assumptions. So far there have been few if any systematic comparisons of the results of applying different semi-analytic models to the same merger tree infrastructure, or conversely, of applying the same SAM to merger trees obtained by applying different algorithms  to the same simulation. This would be very illuminating, clarifying some of the systematic uncertainties in SAM results, as well as their relation to corresponding astrophysical uncertainties in  hydrodynamical simulations. For galaxy clustering in wide-field surveys, there is some reason to be optimistic, however, since physical uncertainties in the star formation/feedback modelling may not be a dominant source of uncertainty, at least on large scales. Some superficial evidence for this comes from the fact that the clustering predictions of our MTNG SAM and our MTNG hydro simulation are very close, and in comparatively good agreement with low-redshift data; see the companion paper by \citet{Bose2022}. We also emphasize a central point of the present paper, namely that certain common approximations, such as constructing a lightcone from a small number of snapshots at fixed times, can introduce errors in clustering predictions that are not small given the ambitious accuracy goals of upcoming large surveys. Unlike the physics uncertainties, these errors can easily be eliminated by adopting continuous lightcone modelling based on sufficiently large underlying simulations.

Finally, we note that our new version of {\small L-GALAXIES} has an improved parallelization approach, based on a central scheduler that eliminates work-load imbalances when the code works in parallel on a given input set of trees. This input set could consist of a suitably chosen subset of all trees from a simulation, allowing the impact of parameter variations on galaxy formation and clustering to be explored very quickly. \citet{Henriques2015} and \cite{vanDaalen2016} used this approach to find optimum values for the free parameters of the model based on an MCMC method. We plan to use similar methods in forthcoming work to improve the physical modelling of {\small L-GALAXIES}, and in particular to allow it to match more precisely, if desired, new observational datasets on galaxy properties and clustering, or the results of full hydrodynamical simulations such as our MTNG740 simulation.

\section*{Acknowledgments}

We thank the referee for an insightful and constructive report that helped to improve the manuscript. The authors gratefully acknowledge the Gauss Centre for Supercomputing (GCS) for providing computing time on the GCS Supercomputer SuperMUC-NG at the Leibniz Supercomputing Centre (LRZ) in Garching, Germany, under project pn34mo. This work used the DiRAC@Durham facility managed by the Institute for Computational Cosmology on behalf of the STFC DiRAC HPC Facility, with equipment funded by BEIS capital funding via STFC capital grants ST/K00042X/1, ST/P002293/1, ST/R002371/1 and ST/S002502/1, Durham University and STFC operations grant ST/R000832/1. CH-A acknowledges support from the Excellence Cluster ORIGINS which is funded by the Deutsche Forschungsgemeinschaft (DFG, German Research Foundation) under Germany’s Excellence Strategy – EXC-2094 – 390783311. VS and LH acknowledge support by the Simons Collaboration on “Learning the Universe”. LH is supported by NSF grant AST-1815978.  SB is supported by the UK Research and Innovation (UKRI) Future Leaders Fellowship [grant number MR/V023381/1].  

\section*{Data Availability}

The data underlying this article will be shared upon reasonable request to the corresponding authors. All simulation data of the MillenniumTNG project are foreseen to be made fully publicly available in 2024.


\appendix

\renewcommand{\thefigure}{A1}
\begin{figure*}
    \centering
 \includegraphics[width=0.95\textwidth]{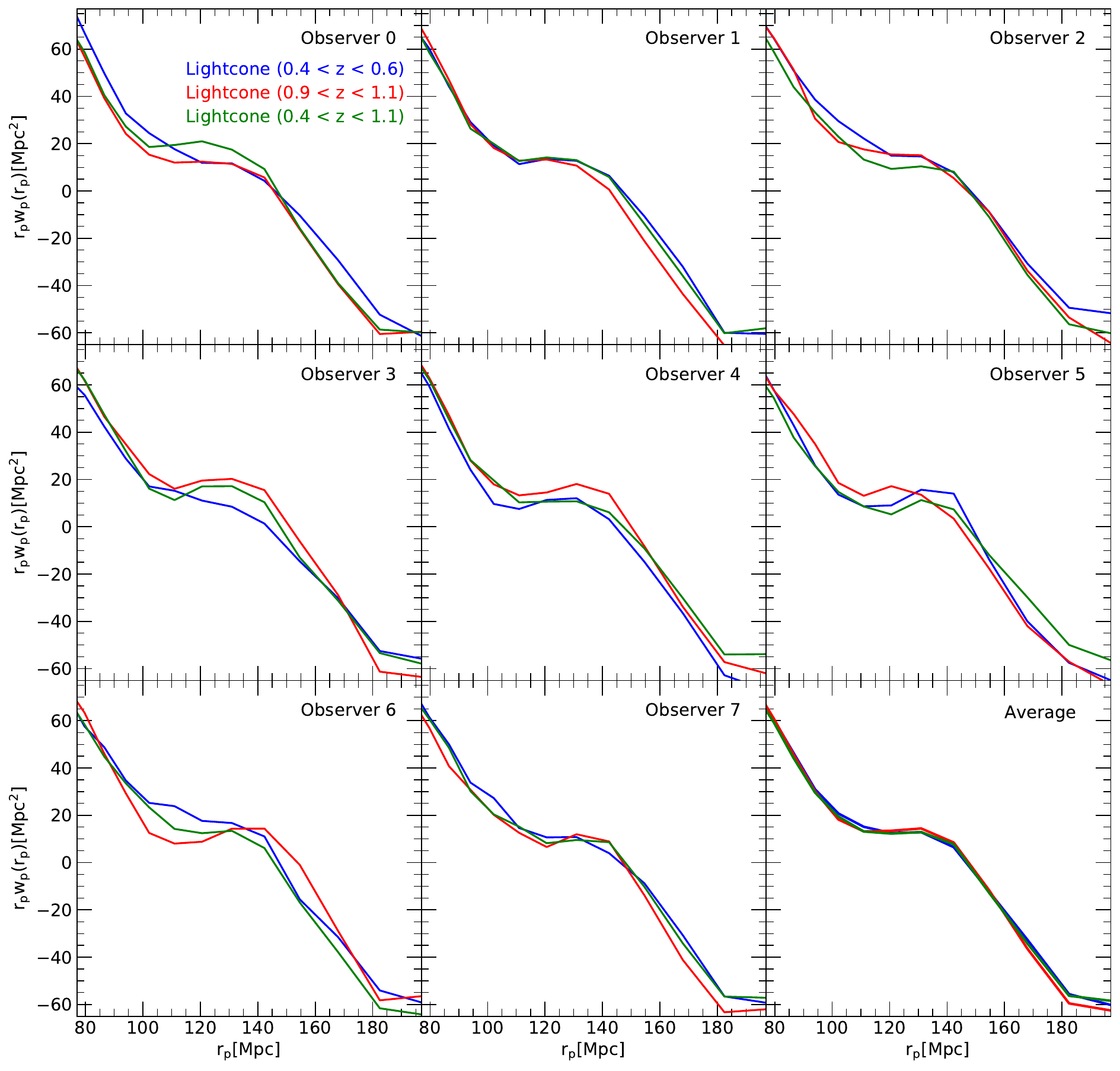}
    \caption{Projected two-point galaxy correlation functions measured for 8 different lightcones constructed for randomly chosen observer positions by translating the whole universe constructed from the $M_\star > 10^{10}\,{\rm M}_\odot$ galaxy distribution of the $z=0.5$ snapshot of the MTNG740-A simulation, neglecting evolution and redshift space distortions. The first 8 panels show the results for redshift shells equal to $0.4 < z < 0.6$, $0.9<z<1.1$ and $0.4 < z < 1.1$, as labelled, while the bottom right panel gives their averages over all 8 realisations. Even though these lightcones are unaffected by galaxy evolution and redshift space distortions, the estimated correlations vary significantly at the BAO scale as a result of the non-uniform coverage of the lightcone shells by the periodic simulation box. Averaging over an ensemble of observer positions (bottom right) eliminates this effect.} 
    \label{fig:twoPointDifferentObservers}
\end{figure*}

\renewcommand{\thefigure}{A2}
\begin{figure}
    \centering    
\includegraphics[width=0.49\textwidth]{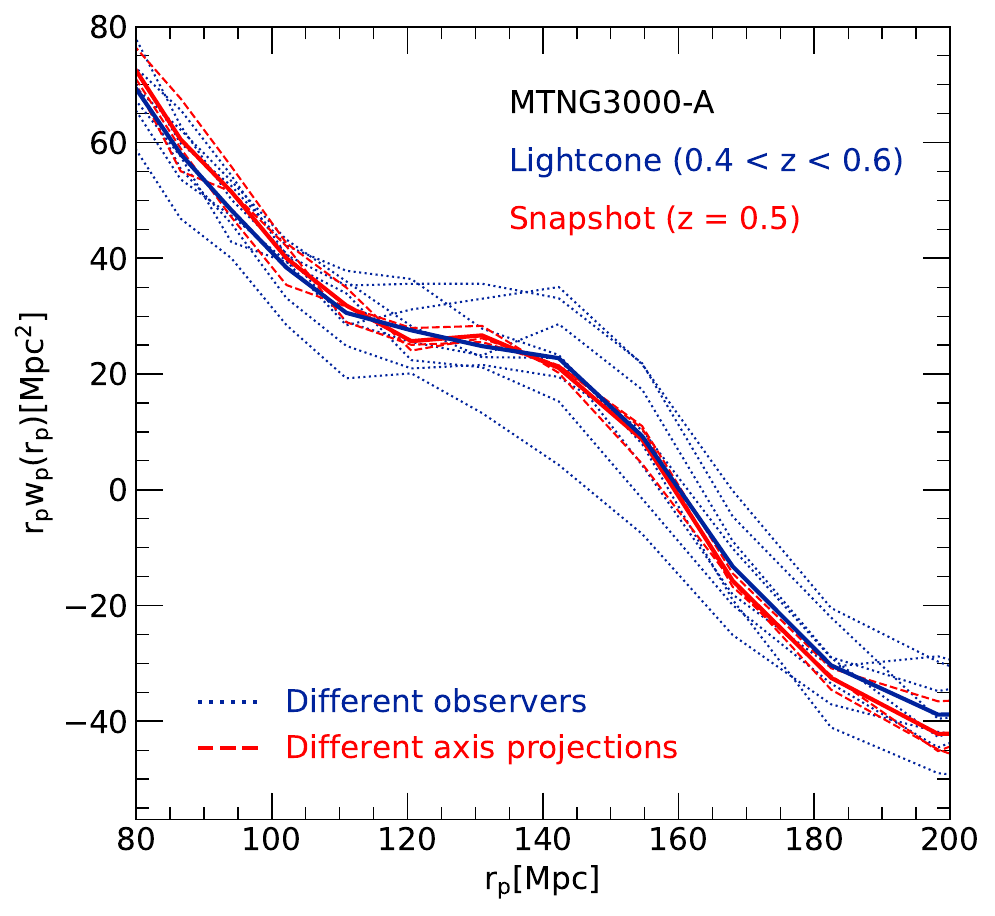}
    \caption{Similar to Fig.~\ref{fig:twoPointDifferentObservers}, but now using the MTNG3000-A simulation as a base. Here we show 8 measurement results for different observer locations for the redshift shell $0.4 < z < 0.6$ as dashed lines, together with their average as a solid line. We also include the measurement obtained for the snapshot directly, computed as the average of projections along the three principal coordinate directions, which are also shown separately (dashed). For this big box, the snapshot results show no sign of residual anisotropy of the correlation function at the BAO scale, and the result for each projection direction agrees individually with the ensemble average of the lightcone shells. However, the lightcone shells themselves still show significant cosmic variance effects as a result of a non-uniform coverage of the lightcone shell with the snapshot volume, and the many independent large-scale modes contained in the $3000\,{\rm  Mpc}$ box. }
    \label{fig:DifferentObserversMTNG3000}
\end{figure}

\bibliographystyle{mnras}
\bibliography{MTNG_bib}


\section{Cosmic variance effects for the clustering signal on lightcone shells}
\label{appendix:lightconeclustering}

The results of Figure~\ref{fig:twoPoint_zoom} for the projected galaxy two-point correlation function show systematic clustering amplitude differences between the two thin and the thick lightcone shells that are larger than the statistical uncertainties of the measurements. This is despite the fact that the results for the snapshots at $z=0.5$ and $z=1.0$ agree quite well, suggesting that the disagreement cannot be blamed, for example, on significant temporal evolution of the galaxy population over this redshift range.  Here we demonstrate that this is primarily an effect of cosmic variance due to the fact that the redshift shells are non-uniformly covered by the simulation box, i.e.~that not all regions of the box are mapped an equal number of times onto the lightcone shell.

To demonstrate this explicitly, we have taken the $z=0.5$ snapshot of the MTNG740-A simulation and have used it to tessellate the comoving backwards lightcone, using periodic replication. This corresponds to how the lightcone was constructed, except that the galaxy population does not evolve, by construction, and we have also, for clarity, omitted redshift space distortions. 

Next we have chosen 8 random locations for the origin of the lightcone inside the box, and have measured the projected correlation function just like in Fig.~\ref{fig:twoPoint_zoom} for the three redshift shells $0.4 < z < 0.6$, $0.9<z< 1.1$, and $0.4 < z < 1.1$. We show the outcome in Fig.~\ref{fig:twoPointDifferentObservers}. Interestingly, the results for the lightcone shells vary substantially with observer location. The bottom right panel shows the averages over these 8 realisations. While these averages agree with each other very well, for  individual observers significant offsets from the mean can occur. One might have expected that better agreement -- after all, the box volume of $(500\,h^{-1}{\rm Mpc})^3$ fits already $\sim 79.5$ times into the comoving volume of the nearer shell.\footnote{The comoving distance out to $z=0.4$ is  $1083.35\,h^{-1}{\rm Mpc}$, and to $z=0.6$ it is $1538.94\,h^{-1}{\rm Mpc}$.} However, it is important to realise that the coverage of the shell does not involve randomly shifted copies of the fundamental box, rather these copies are correlated through the periodic replication condition. As a result, not every point in the box appears ~79 times in the shell. In fact, some areas of the box will appear 70 times in the lightcone shell, while others enter more than 95 times. This non-uniformity of the coverage makes the lightcone shell results vary with observer position, and for this reason, one also cannot expect an individual lightcone shell to reproduce the snapshot result perfectly.

It is interesting to look at this also for the big box. In Fig.~\ref{fig:DifferentObserversMTNG3000} we show measurements for the $z=0.5$ snapshot of the MTNG3000-A run (results projected along the three coordinate axes are shown as dashed lines, and their average as a solid line), and for artificial lightcones constructed from it for 8 observer positions as in Figure~\ref{fig:twoPointDifferentObservers}, (dotted lines). The average of the lightcone results reproduces the snapshot result essentially perfectly, but there is an even a larger variation between the individual estimates than in Figure~\ref{fig:twoPointDifferentObservers}. Again, this may perhaps seem surprising at first, but the underlying effect is the same as above. For the big box, our low-redshift lightcone shell now has 1.17 times the volume of the box. But when we look at how the shell is covered by the box in detail, we realise that this is again fairly non-uniform. More than 20\% of the box-volume is not  mapped into the lightcone shell at all, while about five percent of points in the box appear three times, and of order one percent even four times. This non-uniform coverage perturbs the result from what one gets for uniform analysis of the full snapshot volume. However, averaging  over many observer positions recovers the snapshot result. Note also that the snapshot result is robust with respect to the coordinate axis used for projecting the correlation function, showing that the box volume is large enough to eliminate cosmic variance effects at the snapshot level, whereas they are still present for lightcone shells. Mitigating cosmic variance effects in simulated lightcone measurements of the BAO region and beyond calls for a large comoving simulation volume, but from Earth we can observe only a single past lightcone with opening angle significantly smaller then $4\pi$, so observational surveys of large-scale structure will always have cosmic variance fluctuations at least as large as those between the dotted curves in Figure~\ref{fig:DifferentObserversMTNG3000}.

\section{Speed of the semi-analytic code}
\label{appendix:speed}

For many reasons, high execution speed is, of course, very desirable for the semi-analytic code. In order to quantify realistically the performance difference between our new version of {\small L-GALAXIES} and the older one of \citet{Henriques2015} we have re-processed the original Millennium simulation (which has 64 stored snapshots) with the {\small GADGET-4} code to produce merger trees in the new, modern format (using the {\small SUBFIND-HBT} algorithm). This allowed the new {\small L-GALAXIES} code  applied to Millennium-trees in the modern format, to be compared with the old {\small L-GALAXIES} code applied to Millennium-trees in the old format, using the same computer hardware.
 
In order to produce the $\sim 16.5$~million galaxies that the models predict at $z=0$ in the Millennium simulation\footnote{We have based this comparison on producing results at $z=0$ only, in both cases with a tracking of the star formation history and a photometry computation in 5 bands.}, the old code took 62 minutes on one 40-core node of MPA's local `Freya' compute cluster (Intel Xeon 6138 CPUs). In contrast, the new code takes 30 minutes. There is thus a speed-up of slightly more than a factor of 2 resulting from the various code refinements and efficiency optimisations we have implemented, even though our new algorithms also introduce a few moderately costly operations that were not present before, such as a distance computation to all conceivable progenitor galaxies for a given halo centre to select the closest galaxy as the most likely central galaxy of the halo. Of course, for the MTNG simulations, the considerably higher snapshot frequency and higher mass resolution compared to the Millennium simulation will significantly reduce the rate at which semi-analytic catalogues can be computed. In fact, to produce the $\sim 1.5\times 10^9$ galaxies in MTNG3000, our code needs about 30 hours on 16 compute nodes, which is about an order of magnitude more CPU time per galaxy compared to simply scaling up the computational time for the comparatively low-resolution Millennium simulation. This computational cost for a full galaxy catalogue for MTNG3000 corresponds to about 19 thousand core hours. While this represents a non-trivial computational effort it is still more than a few times $10^4$ less effort than computing the underlying dark matter simulation in the first place.



\bsp	
\label{lastpage}
\end{document}